\documentclass[aps,prd,superscriptaddress,showpacs,preprintnumbers]{revtex4}
\usepackage{graphicx,color}
\usepackage{bm}
\newcommand{\be}{\begin{equation}}
\newcommand{\ee}{\end{equation}}
\newcommand{\bea}{\begin{eqnarray}}
\newcommand{\eea}{\end{eqnarray}}

\newcommand{\bfk}{\mbox{\boldmath $k$}}

\def\kt{k_\perp}
\newcommand{\bfp}{\mbox{\boldmath $p$}}

\newcommand{\bfP}{\mbox{\boldmath $P$}}
\newcommand{\bfS}{\mbox{\boldmath $S$}}
\newcommand{\bfs}{\mbox{\boldmath $s$}}

\newcommand{\pup}{p^\uparrow}

\newcommand{\qup}{q^\uparrow}

\newcommand{\la}{\lambda}

\newcommand{\spa}[3]{\left\langle#1\,#3\right\rangle}
\newcommand{\spb}[3]{\left[#1\,#3\right]}

\def\lsim{\mathrel{\rlap{\lower4pt\hbox{\hskip1pt$\sim$}}\raise1pt\hbox{$<$}}}
\def\gsim{\mathrel{\rlap{\lower4pt\hbox{\hskip1pt$\sim$}}\raise1pt\hbox{$>$}}}
\def\nostrocostruttino#1\over#2{\mathrel{\mathop{\kern 0pt \rlap
{\hbox{$#1$}}} \hbox{\kern-.135em $#2$}}}
\def\sumint{\nostrocostruttino \sum \over {\displaystyle\int}}
\begin{document}
\title{Single spin asymmetries in $\ell \, p \to h \, X$ processes: a test of
factorization}

\author{M.~Anselmino}
\affiliation{Dipartimento di Fisica Teorica, Universit\`a di Torino,
             Via P.~Giuria 1, I-10125 Torino, Italy}
\affiliation{INFN, Sezione di Torino, Via P.~Giuria 1, I-10125 Torino, Italy}
\author{M.~Boglione}
\affiliation{Dipartimento di Fisica Teorica, Universit\`a di Torino,
             Via P.~Giuria 1, I-10125 Torino, Italy}
\affiliation{INFN, Sezione di Torino, Via P.~Giuria 1, I-10125 Torino, Italy}
\author{U.~D'Alesio}
\affiliation{Dipartimento di Fisica, Universit\`a di Cagliari,
             I-09042 Monserrato (CA), Italy}
\affiliation{INFN, Sezione di Cagliari,
             C.P.~170, I-09042 Monserrato (CA), Italy}
\author{S.~Melis}
\affiliation{Dipartimento di Scienze e Tecnologie Avanzate, Universit\`a del Piemonte Orientale, \\
             Viale T.~Michel 11, I-15121 Alessandria, Italy}
\affiliation{INFN, Sezione di Torino, Via P.~Giuria 1, I-10125 Torino, Italy}
\author{F.~Murgia}
\affiliation{INFN, Sezione di Cagliari,
             C.P.~170, I-09042 Monserrato (CA), Italy}
\author{A.~Prokudin}
\altaffiliation[ Present address: ]{Jefferson Laboratory, 12000 Jefferson Avenue,
                Newport News, VA 23606}
\affiliation{Dipartimento di Fisica Teorica, Universit\`a di Torino,
             Via P.~Giuria 1, I-10125 Torino, Italy}
\affiliation{INFN, Sezione di Torino, Via P.~Giuria 1, I-10125 Torino, Italy}

\date{\today}

\begin{abstract}
Predictions for the transverse single spin asymmetry (SSA), $A_N$, are given
for the inclusive processes $\ell \, \pup \to h \, X$ and
$\ell \, \pup \to {\rm jet} + X$, which could be
measured in operating or future experiments. These estimates are based on
the Sivers distributions and the Collins fragmentation functions which fit
the azimuthal asymmetries measured in semi-inclusive deep inelastic
scattering (SIDIS) processes ($\ell \, \pup \to \ell' \, h \, X$). The
factorization in terms of transverse momentum dependent distribution and
fragmentation functions (TMD factorization) -- which supplies the theoretical
framework in which SIDIS azimuthal asymmetries are analyzed -- is assumed to
hold also for the $\ell \, p \to h \, X$ inclusive process at large $P_T$.
A measurement of $A_N$ would then provide a direct test of the validity of the TMD
factorization in this case and would have important consequences for the study
and understanding of SSAs in $ p \, \pup \to h \, X$ processes.

\end{abstract}

\pacs{13.88.+e, 13.60.-r, 13.85.Ni}

\maketitle

\section{\label{Intro} Introduction}

Transverse single spin asymmetries (SSAs) in semi-inclusive deep inelastic
scattering (SIDIS), $\ell \, N \to \ell' \, h \, X$, have been measured by
HERMES~\cite{Airapetian:2004tw,Diefenthaler:2007rj,Pappalardo:2008zz,Airapetian:2009ti} and
COMPASS~\cite{Alexakhin:2005iw,Ageev:2006da,
Martin:2007au,Alekseev:2008dn}. A large amount of data is still
being analyzed by these Collaborations and new results are expected
soon from the JLab experiments at 6 GeV. A rich program focused on
azimuthal asymmetries, as a way of probing the internal nucleon
structure, is planned for JLab operating at an upgraded energy of 12
GeV and for the future electron-ion (EIC) or electron-nucleon (ENC)
colliders, which are under active consideration within the hadron
physics scientific community (see e.g.~Ref.~\cite{DeRoeck:2009af} for a short up-to-date overview).

These SIDIS SSAs are interpreted and discussed in terms of
unintegrated, transverse momentum dependent, distribution and
fragmentation functions (shortly, TMDs). In particular the Sivers
distributions~\cite{Sivers:1989cc, Sivers:1990fh} and the Collins
fragmentation functions~\cite{Collins:1992kk} have been
extracted~\cite{Anselmino:2005ea,Anselmino:2008sga,
Vogelsang:2005cs,Collins:2005ie,Anselmino:2005an,Efremov:2006qm}
from SIDIS data, and, thanks to complementary information from Belle
on the Collins function~\cite{Abe:2005zx,Seidl:2008xc}, a first
extraction of the transversity distribution has been
possible~\cite{Anselmino:2007fs,Anselmino:2008jk}.

All these analyses have been performed in the $\gamma^* - p$ c.m.~frame,
within a QCD factorization scheme, according to which the SIDIS cross
section is written as a convolution of TMDs and elementary interactions:
\be
{\rm d}\sigma^{\ell p \to \ell' h X} = \sum_q \hat f_{q/p}(x, \bfk_\perp; Q^2)
\otimes {\rm d}\hat\sigma^{\ell q \to \ell q} \otimes
\hat D_{h/q}(z, \bfp_\perp; Q^2) \>,
\ee
where $\bfk_\perp$ and $\bfp_\perp$ are, respectively, the
transverse momentum of the quark in the proton and of the final
hadron with respect to the fragmenting quark. At order $k_\perp/Q$
the observed transverse momentum, $\bm{P}_T$, of the hadron is given by
\be
\bm{P}_T = \bfk_\perp + z \, \bfp_\perp  \>.
\ee
There is a general consensus~\cite{Collins:2002kn,Collins:2004nx,Ji:2004wu,
Ji:2004xq,Bacchetta:2008xw} that such a scheme holds in the kinematical
region defined by
\be
P_T \simeq k_\perp \simeq \Lambda_{\rm QCD} \ll Q \>.
\ee
The presence of the two scales, small $P_T$ and large $Q$, allows to
identify the contribution from the unintegrated partonic distribution
($P_T \simeq \kt$), while remaining in the region of validity of the QCD
parton model. At larger values of $P_T$ other mechanisms, like quark-gluon
correlations and higher order pQCD contributions become
important~\cite{Ji:2006br,Anselmino:2006rv,Bacchetta:2008xw}. A
similar situation~\cite{Collins:1984kg,Collins:2004nx,Ji:2004xq,Anselmino:2002pd,Ji:2006ub,
Ji:2006vf,Arnold:2008kf,Anselmino:2009st} holds for Drell-Yan processes,
$A B \to \ell^+ \ell^- X$, where the two scales are the small transverse
momentum, $q_T$, and the large invariant mass, $M$, of the dilepton pair.

The situation is not so clear for processes in which only one large
scale is detected, like the inclusive production, at large $P_T$, of
a single particle in hadronic interactions, $A B \to C X$. However,
the most striking and large SSAs have
been~\cite{Adams:1991rw,Adams:1991cs,Adams:1991ru,Bravar:1996ki} and keep being
measured~\cite{Adams:2003fx,Adler:2005in,Abelev:2008qb,Arsene:2008mi,Aidala:2008qj}
in these cases. The TMD factorization for these processes was first
suggested in Refs.~\cite{Sivers:1989cc,Sivers:1990fh} and adopted in
Refs.~\cite{Anselmino:1994tv, Anselmino:1998yz, Anselmino:1999pw} to
explain the large single spin asymmetries observed by the E704
Collaboration~\cite{Adams:1991cs,Bravar:1996ki}. The same approach
led to successful predictions~\cite{D'Alesio:2004up,D'Alesio:2007jt}
for the values of $A_N$ measured at RHIC~\cite{Nogach:2006gm}.

Alternative approaches to explain the origin of SSAs, linking collinear
partonic dynamics to higher-twist quark-gluon correlations, were originally
proposed in Refs.~\cite{Efremov:1981sh,Efremov:1984ip,Qiu:1991pp,
Qiu:1998ia,Kanazawa:2000hz} and phenomenologically adopted in
Refs.~\cite{Kouvaris:2006zy,Eguchi:2006qz,Eguchi:2006mc,Koike:2009ge}.
These two approaches, the TMD factorization and the higher-twist
correlations, have been shown to be somewhat
related~\cite{Boer:2003cm,Yuan:2009dw} and consistent with each
other~\cite{Ji:2006ub,Ji:2006vf,Koike:2007dg}.

However, a definite proof of the validity of the TMD factorization for
hadronic inclusive processes with one large scale only is still lacking.
Due to this, the study of dijet production at large $P_T$ in hadronic
processes was proposed~\cite{Boer:2003tx,Bomhof:2004aw,Bacchetta:2005rm,
Bomhof:2007su}, where the second small scale is the total $q_T$
of the two jets, which is of the order of the intrinsic partonic momentum
$k_\perp$. This approach leads to a modified TMD factorization approach,
with the inclusion in the elementary processes of gauge link color
factors~\cite{Bomhof:2006dp,Bacchetta:2007sz,Ratcliffe:2007ye}.

In this paper we propose a phenomenological test of the validity of
the TMD factorization in cases in which only one large scale is
detected, by considering SSAs for the $\ell \, \pup \to h \, X$
process, with the detection, in the lepton-proton c.m.~frame, of a
single large $P_T$ final particle, typically a pion. The final lepton
is not observed; notice, however, that a large value of $P_T$ implies,
at leading perturbative order, large values of $Q^2$. Such a measurement
is the exact analogue of the SSAs observed in the $p \, \pup \to h \, X$
processes, the well known and large left-right asymmetries
$A_N$~\cite{Adams:1991rw,Adams:1991cs,Adams:1991ru,Bravar:1996ki,
Adams:2003fx,Adler:2005in,Abelev:2008qb,Arsene:2008mi,Aidala:2008qj}. We compute
these SSAs assuming the TMD factorization and using the relevant
TMDs (Sivers and Collins functions) as extracted from SIDIS data.

Such a choice is natural for the  Collins function, which is expected to
be universal~\cite{Meissner:2008yf,Yuan:2008yv}. The Sivers distribution, instead, is
expected to be process dependent as it is originated by final (or initial,
depending on the process considered) state interactions, which also
model the gauge links necessary for its correct gauge invariant
definition~\cite{Brodsky:2002cx,Collins:2002kn,Brodsky:2002rv}. However, these final
state interactions should be the same in usual SIDIS processes and in the process
considered here.

A similar idea of computing left-right asymmetries in SIDIS processes,
although with different motivations and still demanding the
observation of the final lepton, has been discussed in
Ref.~\cite{She:2008tu}. A first simplified study of $A_N$ in $\ell
\, \pup \to h \, X$ processes was performed in
Ref.~\cite{Anselmino:1999gd}. The process was also considered in
Refs.~\cite{Koike:2002ti,Koike:2002gm} in the framework of collinear
factorization with twist-three correlation functions, obtaining
anomalously large asymmetries with a sign opposite to that of the
corresponding asymmetries in $p \, p$ processes.

The plan of the paper is the following: in Section~\ref{For} we
present the formalism for the study of SSAs in a TMD approach for
both the $\pup \ell \to h \,X$ and the $\pup \ell \to {\rm jet\/} +
X$ processes; in Section~\ref{Mod} we show our numerical estimates
of the contributions of the Sivers and Collins effects to $A_N$,
based on the present knowledge of TMDs, for several different
kinematical setups and discuss their phenomenological aspects;
finally, in Section~\ref{comm} we give some comments and
conclusions. Technical details on the full noncollinear kinematics
are given in Appendix~\ref{kin}, while the calculation of the
helicity amplitudes is worked out in Appendix~\ref{spinors}. The
complete expression of $A_N$ for the process $\pup\ell \to h\, X$,
including all TMD contributions at leading twist, can be found in
Appendix~\ref{aut}.
\section{\label{For} Formalism}
\subsection{Large $P_T$ hadron production}
We propose to study single spin asymmetries for the the process
$\pup \ell \to h \, X$ in close analogy to the study of the SSAs for
the process $\pup p \to h \, X$, assuming the validity of the TMD
factorization. The cross section for this process can then be
written as a particular case of the general treatment, in a
factorized scheme, of the $(A,S_A) + (B,S_B) \to C + X$ large $P_T$
inclusive polarized
process~\cite{Anselmino:1995vq,D'Alesio:2004up,Anselmino:2004ky,Anselmino:2005sh}:
\bea
\frac{E_h \, {\rm d}\sigma^{(p,S) + \ell \to h + X}}
{{\rm d}^{3} \bm{P}_h} &=& \sum_{q,\{\lambda\}}
\int \frac{{\rm d}x \, {\rm d}z}{16 \, \pi^2 x \, z^2  s} \;
{\rm d}^2 \bfk_{\perp} \, {\rm d}^3 \bfp_{\perp}\,
\delta(\bfp_{\perp} \cdot \hat{\bfp}'_q) \, J(p_\perp)
\> \delta(\hat s + \hat t + \hat u)  \nonumber \\
&\times& \left\{ \rho_{\la^{\,}_q,\la^{\prime}_q}^{q/p,S} \,
\hat f_{q/p,S}(x,\bfk_{\perp}) \> \frac12 \>
\hat M_{\la^{\,}_q, \la^{\,}_\ell; \la^{\,}_q, \la^{\,}_\ell} \,
\hat M^*_{\la^{\prime}_q,\la^{\,}_\ell; \la^{\prime}_q, \la^{\,}_\ell} \> \hat
D^{\la^{\,}_h,\la^{\,}_h}_{\la^{\,}_q,\la^{\prime}_q}(z,\bfp_{\perp}) \right\}
\,, \label{gensig}
\eea
which can be shortened, with obvious notations, as:
\be
{\rm d}\sigma^{S} = \sum_{q,\{\lambda\}}
\int \frac{{\rm d}x \, {\rm d}z}{16 \, \pi^2 x \, z^2  s} \;
{\rm d}^2 \bfk_{\perp} \, {\rm d}^3 \bfp_{\perp}\,
\delta(\bfp_{\perp} \cdot \hat{\bfp}'_q) \, J(p_\perp)
\> \delta(\hat s + \hat t + \hat u)
\> \Sigma(S)^{q \ell \to q \ell}(x,z,\bfk_\perp,\bfp_\perp) \,,
\label{sig}
\ee
where $\Sigma(S)$ is the term in curly brackets of Eq.~(\ref{gensig}).

Let us recall the main features of these equations.
\begin{itemize}
\item
We consider the collision of a polarized proton (or, in general, a
nucleon) in a pure transverse spin state $S$ with an unpolarized
lepton, {\it in the proton-lepton center of mass frame}. The proton
$p$ moves along the positive $Z_{\rm cm}$ axis and hadron $h$ is
produced in the $(XZ)_{\rm cm}$ plane. We define as transverse
polarization for the proton the $Y_{\rm cm}$ direction, often using the
notation $\uparrow$ and $\downarrow$ respectively for protons
polarized along or opposite to $Y_{\rm cm}$. The $X_{\rm cm}$ axis is
defined in such a way that a hadron $h$ with $(P_h)_{X_{\rm cm}} > 0$ is
produced {\it to the left} of the incoming proton. The transverse
momentum is denoted as $\bm{P}_T$.  This kinematical configuration is
shown in Fig.~\ref{fig.1}. Results for the case of leptons moving
along the positive $Z_{\rm cm}$ axis ($\ell \, \pup \to h \, X$) will
also be discussed in the paper.
%
\begin{figure*}[t]
\vskip -48pt
\includegraphics[width=15.0truecm]{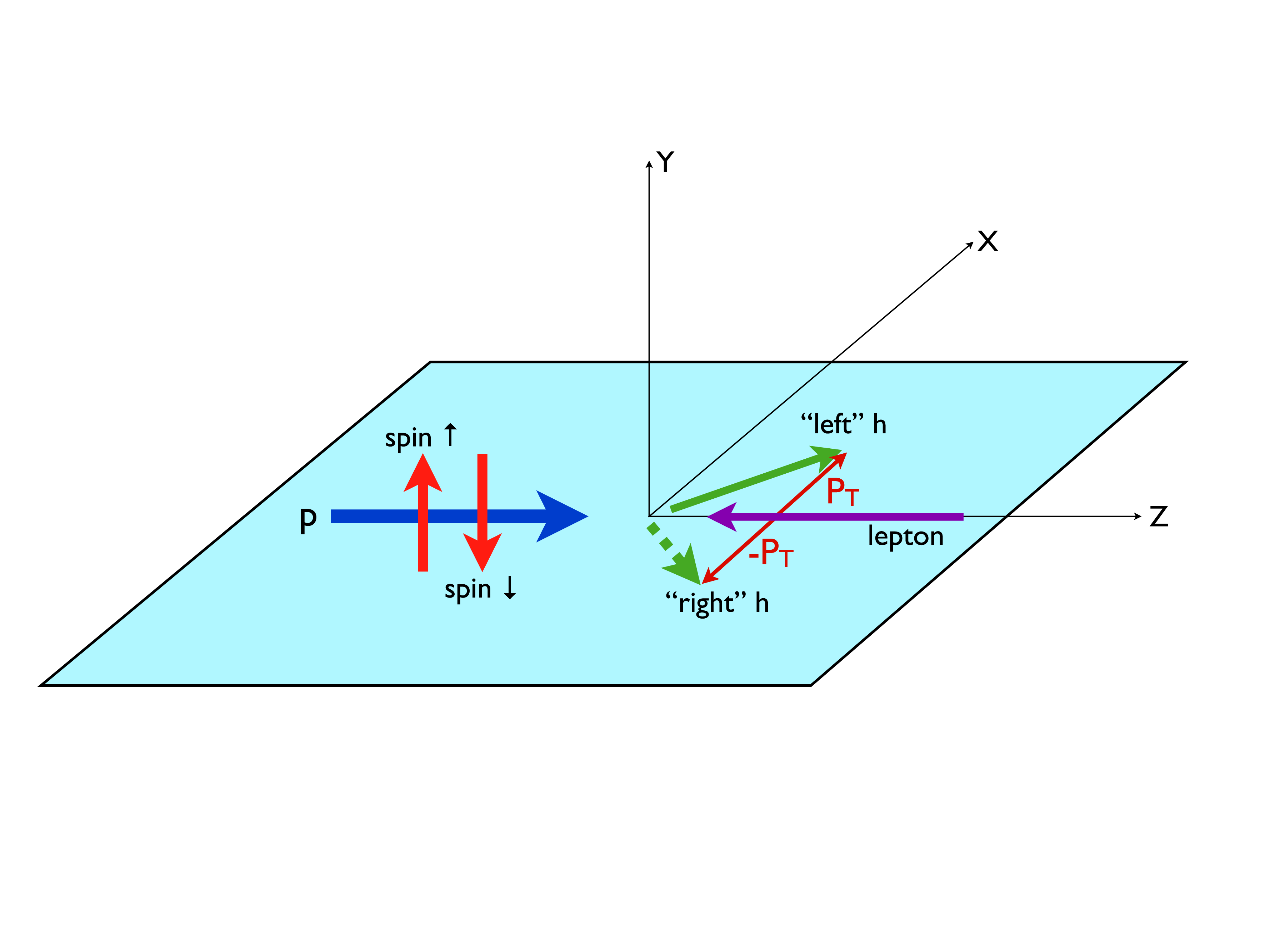}
\vskip -80pt
\caption{\small Kinematical configuration and conventions for the
$\pup \ell \to h \, X$ process.}
\label{fig.1}
\end{figure*}
\item
The notation $\{\lambda\}$ implies a sum over {\it all} helicity indices.
$x$ and $z$ are the usual light-cone momentum fractions, of partons in hadrons
($x$) and hadrons in partons ($z$). $\bfk_{\perp}$ and $\bfp_{\perp}$ are
respectively the transverse momentum of the parton $q$ with respect to its
parent nucleon $p$, and of hadron $h$ with respect to its parent parton $q$.
$\bfp'_q$ is the three-momentum of the final fragmenting parton; it can be
expressed in terms of the integration variables and the observed final
hadron momentum. We consider all partons as massless, neglecting heavy
quark contributions. Full details can be found in Ref.~\cite{Anselmino:2005sh} and
useful expressions are given in Appendix~\ref{kin}.
\item
With massless partons, the function $J$ is given by~\cite{D'Alesio:2004up}
\be\label{jacobian}
J(p_{\perp}) =
\frac{\left( E_h + \displaystyle{\sqrt{\bm{P}_h^{2} - \bfp^2_{\perp }}} \right)^2}
{4(\bm{P}_h^{2} - \bfp^2_{\perp})}
\>\cdot
\ee
In the kinematical regions which we shall consider $J$ is close to 1.
\item
$\rho_{\la^{\,}_q, \la^{\prime}_q}^{q/p,S}$ is the helicity density matrix
of parton $q$ inside the polarized proton $p$, with spin state $S$.
$\hat f_{q/p,S}(x,\bfk_{\perp})$ is the distribution function of the
unpolarized parton $q$ inside the polarized proton $p$. The products
$\rho_{\la^{\,}_q, \la^{\prime}_q}^{q/p,S} \> \hat f_{q/p,S}(x,\bfk_{\perp})$
are directly related to the leading-twist TMDs, with a dependence on $\phi$,
the azimuthal angle of $\bfk_\perp$~\cite{Anselmino:2005sh}.
\item
The $\hat M_{\la^{\,}_q, \la^{\,}_\ell;\la^{\,}_q, \la^{\,}_\ell}$'s are the
helicity amplitudes for the elementary process $q \, \ell \to q \, \ell$,
normalized so that the unpolarized cross section, for a collinear collision,
is given by
\be
\frac{{\rm d}\hat\sigma^{q\ell \to q\ell}}{{\rm d}\hat t} = \frac{1}{16\pi\hat s^2}
\frac{1}{4} \sum_{\la^{\,}_q, \la^{\,}_\ell}
|\hat M_{\la^{\,}_q, \la^{\,}_\ell; \la^{\,}_q, \la^{\,}_\ell}|^2\,.
\label{norm}
\ee
At lowest perturbative order $q \, \ell \to q \, \ell$ is the only elementary
interaction which contributes; notice that, in the presence of parton intrinsic
motion, it is not a planar process in our chosen frame and depends on the
intrinsic momenta, including their phases. Neglecting lepton and quark masses
there are two independent helicity amplitudes:
\bea
\hat M_{++;++}(\hat s, \hat t, \hat u, \bfk_\perp) = \hat M_{--;--}^* &=&
- 8 \, \pi \, e_q \, \alpha \, \frac{\hat s}{\hat t} \> e^{i \varphi_1}
\equiv \hat M_1^0 \> e^{i \varphi_1} \label{ampl1} \\
\hat M_{+-;+-}(\hat s, \hat t, \hat u, \bfk_\perp) = \hat M_{-+;-+}^* &=&
8 \, \pi \, e_q \, \alpha \, \frac{\hat u}{\hat t} \> e^{i \varphi_2}
\equiv \hat M_2^0 \> e^{i \varphi_2}\label{ampl2} \>,
\eea
where $\varphi_{1,2}$ are phases explicitly given in Appendix~\ref{spinors},
Eqs.~(\ref{m++}) and (\ref{m+-}).
\item
$\hat D^{\la^{\,}_h,\la^{\prime}_h}_{\la^{\,}_q,\la^{\prime}_q}(z,\bfp_{\perp})$
is the product of {\it fragmentation amplitudes} for the $q \to h + X$ process
\be
\hat D^{\la^{\,}_h,\la^{\prime}_h}_{\la^{\,}_q,\la^{\prime}_q}
= \> \sumint_{X, \la_{X}} {\hat{\cal D}}_{\la^{\,}_{h},\,\la^{}_X;
\la^{\,}_q} \, {\hat{\cal D}}^*_{\la^{\prime}_h,\,\la^{}_{X}; \la^{\prime}_q}
\, ,
\ee
where the $\sumint_{X, \la_{X}}$ stands for a spin sum and phase space
integration over all undetected particles, considered as a system $X$.
The usual unpolarized fragmentation function  $D_{h/q}(z)$, {\it i.e.}~the
number density of hadrons $h$ resulting from the fragmentation of an
unpolarized parton $q$ and carrying a light-cone momentum fraction $z$,
is given by
\be
D_{h/q}(z) = \frac{1}{2} \sum_{\la^{\,}_q,\la^{\,}_h} \int {\rm d}^2\bfp_{\perp}
\, \hat D^{\la^{\,}_h,\la^{\,}_h}_{\la^{\,}_q,\la^{\,}_q}(z, \bfp_{\perp})
\,. \label{fr}
\ee
We shall only consider the case of spinless final particles ($\la^{\,}_h = 0$),
in particular pions. In general
$\hat D_{\la^{\,}_q,\la^{\prime}_q}(z,\bfp_{\perp})$ depends on the azimuthal
angle of $h$ around the direction of motion of the fragmenting polarized
parton~\cite{Anselmino:2005sh}.
\end{itemize}

We compute the SSA:
\be
A_N = \frac{{\rm d}\sigma^\uparrow(\bm{P}_T) - {\rm d}\sigma^\downarrow(\bm{P}_T)}
           {{\rm d}\sigma^\uparrow(\bm{P}_T) + {\rm d}\sigma^\downarrow(\bm{P}_T)}
    = \frac{{\rm d}\sigma^\uparrow(\bm{P}_T) - {\rm d}\sigma^\uparrow(-\bm{P}_T)}
           {2 \, {\rm d}\sigma^{\rm unp}(\bm{P}_T)} \,, \label{an}
\ee
which can be measured either by looking at the production of hadrons
at a fixed transverse momentum $\bm{P}_T$, changing the incoming proton
polarization from $\uparrow$ to $\downarrow$, or keeping a fixed
proton polarization and looking at the hadron production to the left
and the right of the $Z_{\rm cm}$ axis, see Fig.~\ref{fig.1}. $A_N$ is defined (and computed)
for a proton polarization normal ($N$) to the production plane and a
pure spin state (a pseudo-vector polarization $\bfS_T$ with
$|\bfS_T| = S_T = 1$). For a generic transverse polarization along an
azimuthal direction $\phi_S$ (in our chosen reference frame) and a
polarization $S_T \not= 1$, one has:
\be
A(\phi_S, S_T) = \bfS_T \cdot (\hat{\bfp} \times \hat{\bm{P}}_T) \, A_N =
S_T \sin\phi_S \, A_N \>.
\ee
Notice that if, according to the usual procedure in SIDIS experiments,
one defines
\be
A_{TU}^{\sin\phi_S} \equiv \frac{2}{S_T} \,
\frac{\int \, {\rm d}\phi_S \> [{\rm d}\sigma(\phi_S) - {\rm d}\sigma(\phi_S + \pi)]\> \sin\phi_S}
     {\int \, {\rm d}\phi_S \> [{\rm d}\sigma(\phi_S) + {\rm d}\sigma(\phi_S + \pi)]} \>,
\label{ATU}
\ee
one simply has
\be
A_{TU}^{\sin\phi_S} = A_N \>.
\ee

In order to compute $A_N$, Eq.~(\ref{an}), we need to compute
$[\Sigma(\uparrow) - \Sigma(\downarrow)]$ and $[\Sigma(\uparrow) +
\Sigma(\downarrow)]$, which can be done by performing the helicity
sum in Eqs.~(\ref{gensig}) and (\ref{sig}). As our process is a
simple particular case of $(A,S_A) + (B,S_B) \to C + X$, the result
agrees with Eqs.~(82) and (86) of Ref.~\cite{Anselmino:2005sh},
simplified to the case in which particle $B$ is a point-like lepton
and the elementary interaction has only two independent amplitudes.
Notice that several TMDs appear in the expression for $A_N$;
however, numerical evaluations show that the contribution of the
Sivers effect is the dominant one. A modest contribution is given by
the Collins function (coupled to the transversity distribution),
while another contribution involving $h_{1T}^\perp$ (see
Appendix~\ref{aut}) is totally negligible. Considering only the
Sivers and Collins effects, one has:
\be
A_N =
\frac
{{\displaystyle \sum_{q,\{\lambda\}} \int \frac{{\rm d}x \, {\rm d}z}
{16\,\pi^2 x\,z^2 s}}\;
{\rm d}^2 \bfk_{\perp} \, {\rm d}^3 \bfp_{\perp}\,
\delta(\bfp_{\perp} \cdot \hat{\bfp}'_q) \, J(p_\perp)
\> \delta(\hat s + \hat t + \hat u)
\> [\Sigma(\uparrow) - \Sigma(\downarrow)]^{q \ell \to q \ell}}
{{\displaystyle \sum_{q,\{\lambda\}} \int \frac{{\rm d}x \, {\rm d}z}
{16\,\pi^2 x\,z^2 s}}\;
{\rm d}^2 \bfk_{\perp} \, {\rm d}^3 \bfp_{\perp}\,
\delta(\bfp_{\perp} \cdot \hat{\bfp}'_q) \, J(p_\perp)
\> \delta(\hat s + \hat t + \hat u)
\> [\Sigma(\uparrow) + \Sigma(\downarrow)]^{q \ell \to q \ell}} \>,
\label{anh}
\ee
with
\bea \sum_{\{\lambda\}}\,[\Sigma(\uparrow) - \Sigma(\downarrow)]^{q
\ell \to q \ell} &=& \frac{1}{2} \, \Delta^N\! f_{q/\pup}
(x,k_{\perp}) \cos\phi \, \left[\,|{\hat M}_1^0|^2 + |{\hat
M}_2^0|^2 \right] \,
D_{h/q} (z, p_{\perp}) \label{ds1} \nonumber \\
&+& h_{1q}(x,k_{\perp}) \, \hat M_1^0 \hat M_2^0 \, \Delta^N\!
D_{h/\qup} (z, p_{\perp}) \, \cos(\phi' + \phi_q^h) \label{anhn}
\eea
and (dropping negligible contributions from other TMDs~\cite{Anselmino:2005sh})
\be
\sum_{\{\lambda\}}\,[\Sigma(\uparrow) +
\Sigma(\downarrow)]^{q \ell \to q \ell} =
f_{q/p} (x,k_{\perp}) \,
\left[\,|{\hat M}_1^0|^2 + |{\hat M}_2^0|^2 \right] \,
D_{h/q} (z, p_{\perp}) \>. \label{ss1}
\ee
\begin{itemize}
\item
The first term on the r.h.s.~of Eq.~(\ref{ds1}) shows the
contribution to $A_N$ of the Sivers function $\Delta^N\!
f_{q/\pup}(x,
k_\perp)$~\cite{Sivers:1989cc,Sivers:1990fh,Bacchetta:2004jz},
\bea \Delta \hat f_{q/p,S}(x, \bfk_{\perp}) = \hat f_{q/p,S}(x,
\bfk_{\perp}) - \hat f_{q/p,-S}(x, \bfk_{\perp}) &\equiv& \Delta^N\!
f_{q/\pup}\,(x, k_{\perp}) \>
\bfS_T \cdot (\hat{\bfp} \times \hat{\bfk}_{\perp }) \label{defsivnoi} \\
&=& -2 \, \frac{k_\perp}{M} \, f_{1T}^{\perp q}(x, k_{\perp}) \>
\bfS_T \cdot (\hat{\bfp} \times \hat{\bfk}_{\perp }) \>, \nonumber
\eea
coupled to the unpolarized elementary interaction ($\propto \frac 12
(|\hat{M}_1^0|^2 + |\hat{M}_2^0|^2)$) and the unpolarized fragmentation function
$D_{h/q} (z, p_{\perp})$; the $\cos\phi$ factor arises from the
$\bfS_T \cdot (\hat{\bfp} \times \hat{\bfk}_{\perp })$ factor, the
spin--transverse motion correlation of the Sivers function in the
case of a normal spin direction with $S_T=1$.
\item
The second term on the r.h.s.~of Eq.~(\ref{ds1}) shows the
contribution to $A_N$ of the unintegrated transversity distribution
$h_{1q}(x,k_{\perp})$ coupled to the Collins function $\Delta^N\!
D_{h/\qup} (z, p_{\perp})$~\cite{Collins:1992kk,Bacchetta:2004jz},
\bea \Delta \hat D_{h/q^\uparrow}\,(z, \bfp_{\perp}) = \hat
D_{h/q^\uparrow}\,(z, \bfp_{\perp}) - \hat D_{h/q^\downarrow}\,(z,
\bfp_{\perp}) &\equiv& \Delta^N\! D_{h/\qup}\,(z, p_{\perp}) \>
\bfs_q \cdot (\hat{\bfp}_q^\prime \times \hat{\bfp}_{\perp }) \\
\label{defcolnoi} &=& \frac{2 \, p_\perp}{z \, m_h} H_{1}^{\perp q}(z,
p_{\perp}) \> \bfs_q \cdot (\hat{\bfp}_q^\prime \times
\hat{\bfp}_{\perp}) \>, \nonumber \eea
and to the transverse spin transfer elementary interaction
(${\rm d}\sigma^{\uparrow, \uparrow} - {\rm d}\sigma^{\uparrow, \downarrow}
\propto \hat M_1^0 \hat M_2^0$). The factor $\cos(\phi' + \phi_q^h)$
arises from phases in the $\bfk_\perp$-dependent transversity distribution,
the Collins function and the elementary polarized interaction. $\phi'$ is the
azimuthal angle of the fragmenting quark (with 3-momentum $\bfp^\prime_q$)
and $\phi_q^h$ is the azimuthal angle of $\bfp_\perp$ around the
$\hat{\bfp}^\prime_q$ direction~\cite{Anselmino:2005sh}.
Their expressions in terms of integration and overall variables can be found
in Appendix~\ref{kin}.
\item
The elementary interaction amplitudes are explicitly given in
Eqs.~(\ref{ampl1}) and (\ref{ampl2}). Notice that the elementary
Mandelstam variables $\hat s, \hat t, \hat u$ are computed taking into
account the full kinematics, and thus depend on the transverse momenta.
\item A final issue which needs to be clarified concerns {\it perturbative
QCD corrections}. Our proposed process involves TMDs coupled to lowest order
perturbative interactions and is driven by a large angle elementary
electromagnetic scattering, $q\,\ell \to q\,\ell$. Some QCD effects, like
soft gluon emissions, are taken into account in the TMDs, as the emission of
soft gluons builds up intrinsic partonic motion. Higher order pQCD corrections
due to genuine hard QCD processes, like $q\,\ell \to q\,\ell\,g$ or
$g\,\ell \to q\,\bar q\,\ell$ are not included in our computation of $A_N$.
These contribute at order $\alpha_s$ to the cross section and can be neglected
at large $Q^2$ values;
moreover, one should notice that events induced by these hard
pQCD elementary interactions result in final states with two fragmenting
partons, {\it i.e.}~two jets, and could be experimentally excluded.
However, these pQCD corrections might be of some relevance and difficult
to disentangle at HERMES, COMPASS or JLab energies.
\end{itemize}
\subsection{Large $P_T$ jet production}
We consider also the most interesting case of SSAs for the inclusive process
$\pup \, \ell \to {\rm jet} + X$. Although it is a difficult process to detect
experimentally and might require future higher energy and luminosity machines,
it would certainly give the most direct access to the Sivers effect, as the
lack of any fragmentation mechanism forbids other contributions. Even more
difficult, the observation of both a jet and a final hadron inside the jet
(with a measurement of its transverse momentum $\bfp_\perp$), would allow a
direct detection of the Collins effect~\cite{Yuan:2007nd}.

In the case of the $\pup \, \ell \to {\rm jet} + X$ process, with no
observation of a single final particle, Eq.~(\ref{gensig})
simplifies to:
\bea
\frac{E_{j} \, {\rm d}\sigma^{(p,S) + \ell \to {\rm jet} + X}}
{{\rm d}^{3} \bm{P}_{j}} &=& \sum_{q,\{\lambda\}}
\int \frac{{\rm d}x}{16 \, \pi^2 x \, s} \;
{\rm d}^2 \bfk_{\perp} \> \delta(\hat s + \hat t + \hat u)  \nonumber \\
&\times& \rho_{\la^{\,}_q,\la^{\prime}_q}^{q/p,S} \,
\hat f_{q/p,S}(x,\bfk_{\perp}) \> \frac12 \>
\hat M_{\la^{\,}_q, \la^{\,}_\ell; \la^{\,}_q, \la^{\,}_\ell} \,
\hat M^*_{\la^{\prime}_q,\la^{\,}_\ell; \la^{\prime}_q, \la^{\,}_\ell} \,,
\label{gensigjet}
\eea
while Eq.~(\ref{anh}) becomes:
\be
A_N^{\rm jet} = \frac
{{\displaystyle \sum_{q,\{\lambda\}} \int \frac{{\rm d}x}
{16\,\pi^2 x\,s}}\;
{\rm d}^2 \bfk_{\perp} \> \delta(\hat s + \hat t + \hat u)
\> [\Sigma(\uparrow) - \Sigma(\downarrow)]^{q \ell \to q \ell}_{\rm jet}}
{{\displaystyle \sum_{q,\{\lambda\}} \int \frac{{\rm d}x}
{16\,\pi^2 x\,s}}\;
{\rm d}^2 \bfk_{\perp} \> \delta(\hat s + \hat t + \hat u)
\> [\Sigma(\uparrow) + \Sigma(\downarrow)]^{q \ell \to q \ell}_{\rm jet}}
\> \cdot \label{anjet}
\ee

In this case the kinematics is very simple and is shown
explicitly in appendix~\ref{kin-jet}.
{}For a generic azimuthal direction $\phi_S$ of the transverse
spin $\bfS_T$, the Sivers function, Eq.~(\ref{defsivnoi}), can be
written as:
\bea \Delta^N\! f_{q/\pup}\,(x, k_{\perp}) \> \bfS_T \cdot
(\hat{\bfp} \times \hat{\bfk}_{\perp }) &=& \Delta^N\!
f_{q/\pup}\,(x, k_{\perp}) \left( \sin\phi_S \, \frac{k_\perp^x}
{k_\perp} -\cos\phi_S \, \frac{k_\perp^y} {k_\perp} \right) \nonumber \\
&=& \Delta^N\! f_{q/\pup}\,(x, k_{\perp})\, \sin(\phi_S -\phi) \,,
\label{defsivnoi2} \eea
and the $\Sigma$ kernels in Eq.~(\ref{anjet}) are
\bea
 \sum_{\{\lambda\}}\,[\Sigma(\uparrow) - \Sigma(\downarrow)]^{q
\ell \to q \ell}_{\rm jet} &=& \frac{1}{2} \, \Delta^N\! f_{q/\pup}
(x,k_{\perp}) \, \sin(\phi_S - \phi) \, \left[\,|{\hat M}_1^0|^2 +
|{\hat M}_2^0|^2 \right] \, \label{dsjet} \\
\sum_{\{\lambda\}}\,[\Sigma(\uparrow) +
\Sigma(\downarrow)]^{q \ell \to q \ell}_{\rm jet} &=&
f_{q/p} (x,k_{\perp}) \,
\left[\,|{\hat M}_1^0|^2 + |{\hat M}_2^0|^2 \right] \>. \label{ssjet}
\eea
The elementary amplitudes are the same as given in Eqs.~(\ref{ampl1}) and
(\ref{ampl2}).
\section{\label{Mod} Estimates for $A_N$}
We have computed the SSA, $A_N$, as defined in Eq.~(\ref{an}) or
(\ref{ATU}), for the large $P_T$ production of pions and jets in
$\pup \ell \to h \, X$ and $\pup \, \ell \to {\rm jet} + X$ processes,
according to the expressions given, respectively, in
Eqs.~(\ref{anh})-(\ref{ss1}) and in Eqs.~(\ref{anjet}), (\ref{dsjet}) (with
$\phi_S = \pi/2$), and (\ref{ssjet}).

Analogous results for the case of leptons moving along the $Z_{\rm cm}$ axis,
$\ell\, p^\uparrow\to h\,({\rm jet})+X$, \emph{in the same chosen hadronic frame} (that is,
keeping fixed the definitions of $x_F = 2P_L/\sqrt s$ and of the $\uparrow,\downarrow$
transverse polarization directions) can be easily obtained using
rotational invariance:
\be
A_N^{\,\ell p^\uparrow\to h ({\rm jet})+X}(x_F,\bm{P}_T) \>=\>
 -  A_N^{\,p^\uparrow \ell \to h ({\rm jet})+X}(-x_F,\bm{P}_T)\>.
\label{an-rot-inv}
\ee

We have used the Sivers distributions as parameterized and extracted -- from
SIDIS data -- in Ref.~\cite{Anselmino:2008sga}; even if the Sivers functions,
being related to final state interactions~\cite{Brodsky:2002cx}, are expected
to be process dependent~\cite{Collins:2002kn}, they should be the same in
SIDIS and the (related) processes considered here, which all originate
from the same $q\,\ell \to q\,\ell$ elementary interaction and
subsequent quark fragmentation. Similarly, we have used the
transversity distributions and Collins functions as parameterized and extracted in
Ref.~\cite{Anselmino:2008jk}. The unpolarized parton distribution functions (PDFs)
and  fragmentation functions (FFs) are taken
respectively from Refs.~\cite{Gluck:1998xa} and~\cite{deFlorian:2007aj}.
\begin{figure*}[t!h]
\includegraphics[width=5.truecm,angle=-90]{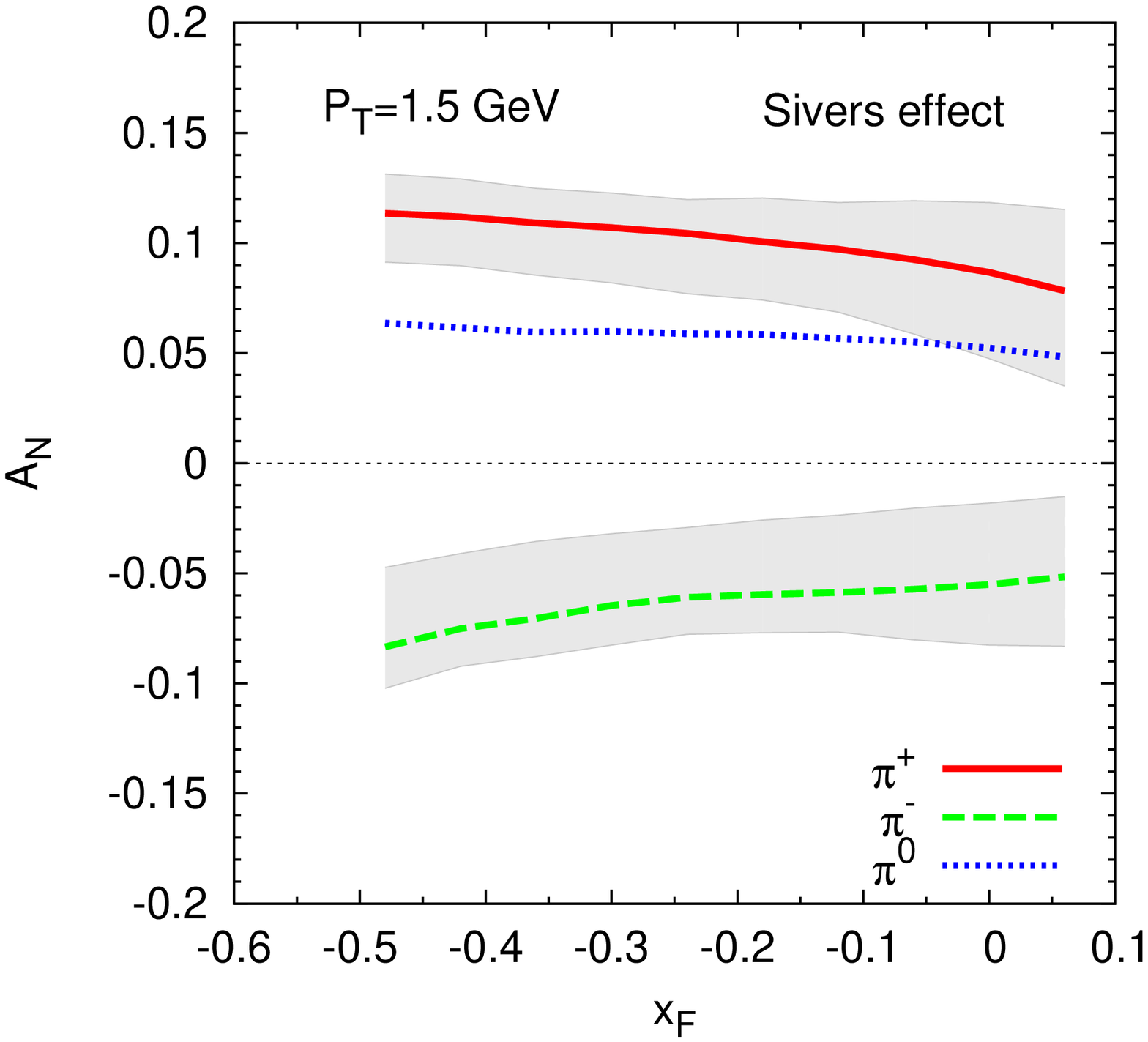}
\includegraphics[width=4.7truecm,angle=-90]{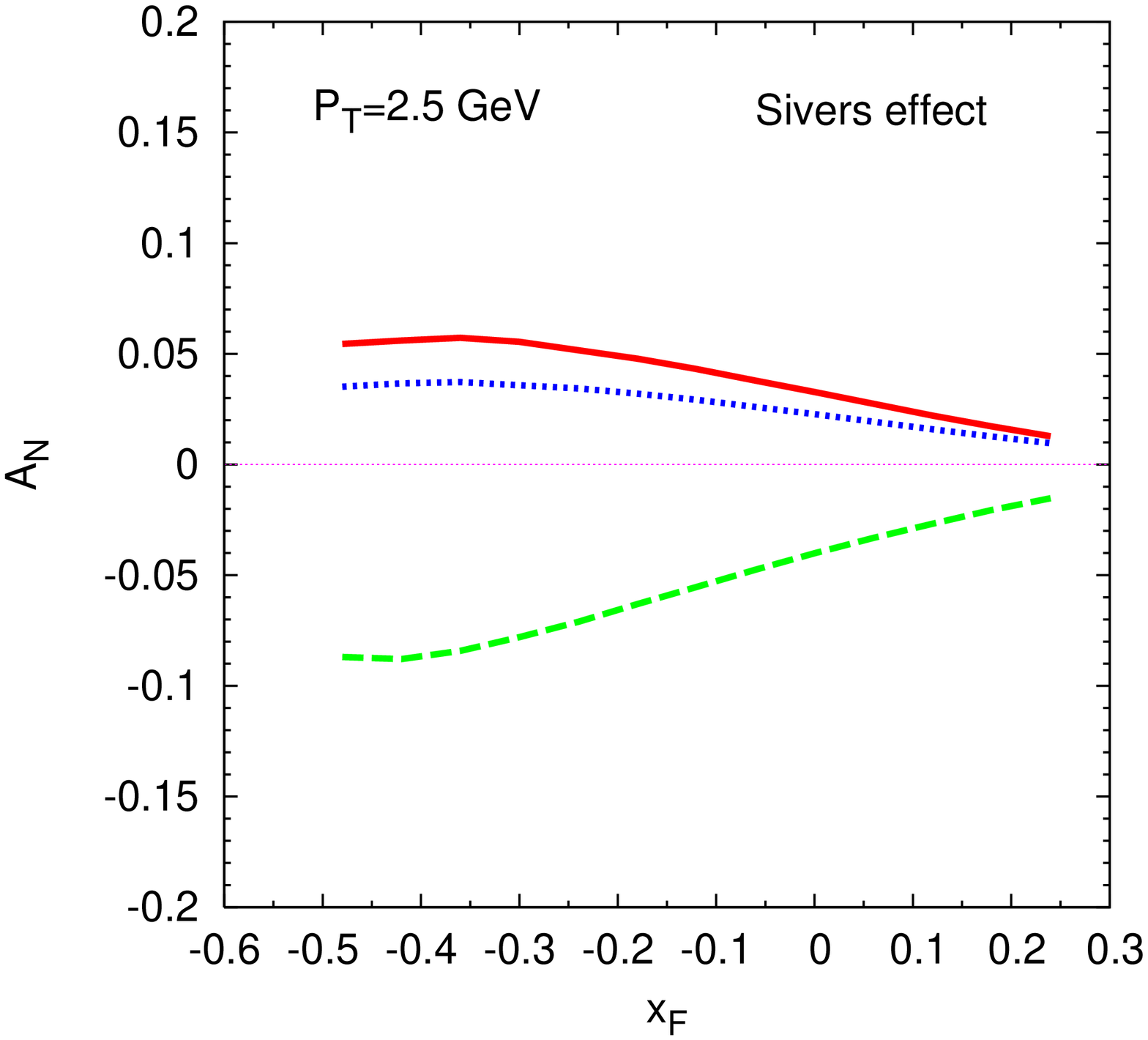}
\includegraphics[width=4.7truecm,angle=-90]{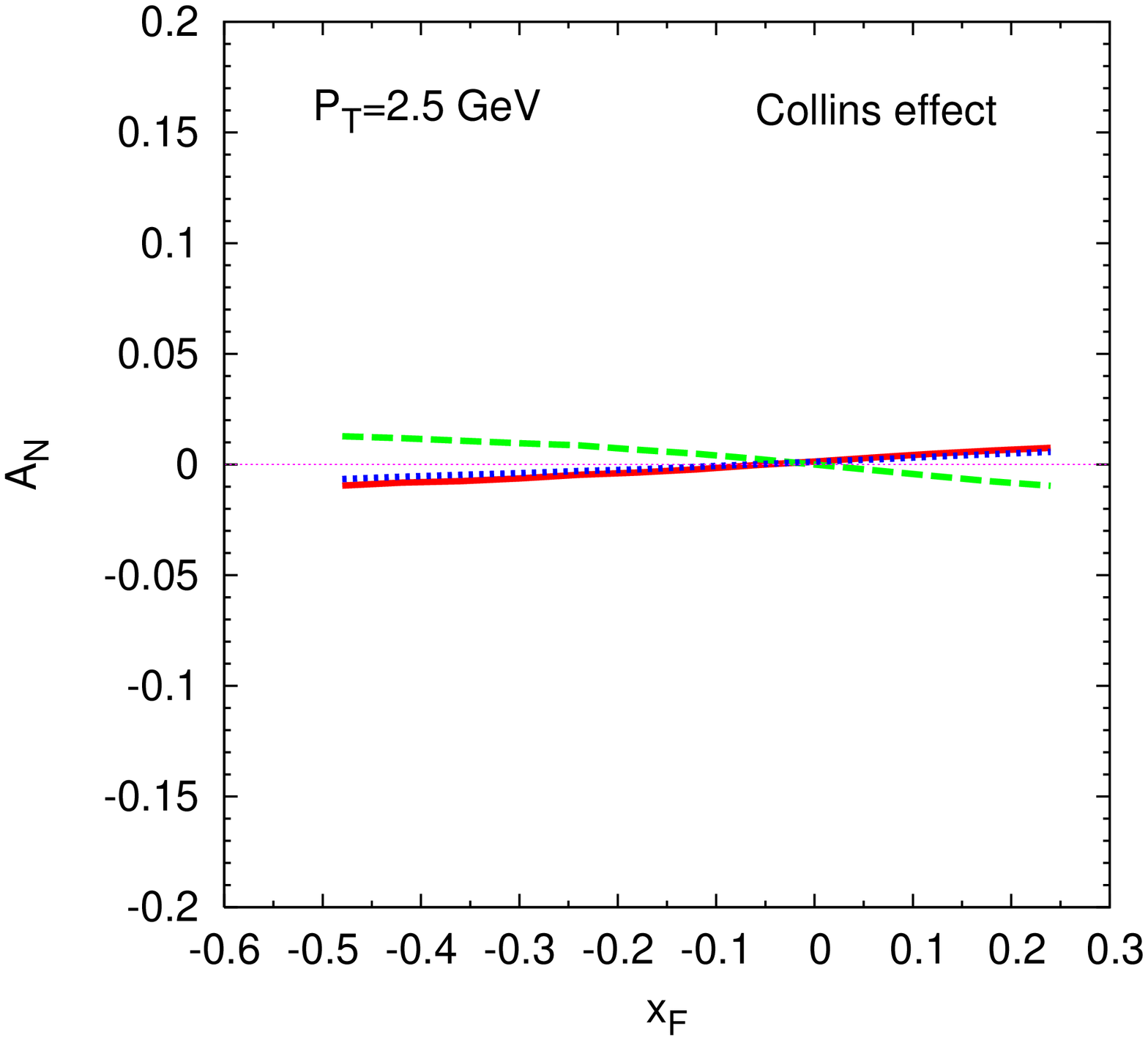}
\caption{Estimates of $A_N$ vs.~$x_F$ for the $\pup \, \ell
\to \pi \, X$ process at HERMES ($\sqrt{s}\simeq 7$ GeV). Left panel: Sivers effect at $P_T =
1.5$~GeV; central panel: Sivers effect at $P_T = 2.5$~GeV; right
panel: Collins effect at $P_T=2.5$ GeV. The computation has been
performed according to Eqs.~(\ref{anh}), (\ref{anhn}) and
(\ref{ss1}) of the text, adopting the Sivers functions of
Ref.~\cite{Anselmino:2008sga} and the transversity and Collins
functions of Ref.~\cite{Anselmino:2008jk}, as extracted from SIDIS
and $e^+e^-$ data, the unpolarized PDFs of Ref.~\cite{Gluck:1998xa}
and the FFs of Ref.~\cite{deFlorian:2007aj}. In the left panel we
also show, for charged pions, the statistical uncertainty bands
coming from the extracted Sivers functions~\cite{Anselmino:2008sga}.}
\label{fig:an-hermes}
\end{figure*}

Our results are given for the kinematical configurations of HERMES,
COMPASS, JLab at 12 GeV, and a hypothetical ENC future machine operating at an energy
$\sqrt s = 50$~GeV. For hadron production, the Sivers and Collins contributions are shown
separately. We plot $A_N$ as a function of $x_F$ at
fixed $P_T$ values; these should be chosen as the hard scale of the
process, ensuring a large momentum transfer in the hard scattering,
say $Q^2 > 1$~GeV$^2$. In collinear cases, at LO, it might suffice to have
$P_T > 1$~GeV; however, with TMD factorization, one has to be more
careful, as $P_T$ might be partially generated by intrinsic
$\bm{k}_\perp$. We have checked that a value of $P_T = 2.5$~GeV
corresponds to a safe $Q^2 > 1$~GeV$^2$ region in the whole range of
$x_F$, while $P_T = 1.5$~GeV implies a safe $Q^2$ region only for
backward production, $x_F \lesssim 0$. We give predictions for these two
values of $P_T$.

Notice also that for positive $x_F$ the minimum of $x$ is given, roughly, by
$x_F$. This implies that for $x_F > 0.2$ -- 0.3 we should employ the
parameterizations of the Sivers and transversity functions in a region
where they are not constrained by SIDIS data. For this reason we will
give our theoretical estimates of $A_N$ only up to $x_F \simeq 0.2$.
On the other hand, for negative $x_F$ the minimum of $x$ is controlled by
the ratio $x_T = 2P_T /\sqrt s$, implying that at moderate c.m.~energies
({\it i.e.}~$\sqrt s \simeq 10$ -- 20 GeV) and $P_T\simeq 1$ -- 2 GeV,
we are sensitive to the valence region of the polarized proton,
{\it i.e.}~the region where the Sivers (and the transversity) functions
reach their maxima.

Let us comment in details our results.
\begin{itemize}
\item
We first stress some aspects peculiar to the $\pup \ell \to h \, X$
process. As in SIDIS processes at leading order accuracy, only one
partonic subprocess, $q \, \ell \to q \, \ell$, is active, with a simple
$1/\hat t^2$ dependence (a much simpler dynamics than in the
$p \, p\to h \,X$ case). However, since the lepton plane is not
identified (we do not require the detection of the outgoing lepton),
one cannot access, separately, the Sivers and the Collins effects.
Nevertheless, in the backward region (w.r.t.~the proton direction)
the variable $|\hat u|$ becomes smaller and so does the partonic spin
transfer cross section $\propto \hat M_1^0 \hat M_2^0$ [see
Eqs.~(\ref{ampl1}) and (\ref{ampl2})], entering the Collins contribution
to $A_N$ [second term on the r.h.s.~of Eq.~(\ref{anhn})]. This implies
a strong {\em dynamical} suppression of the Collins effect (reinforced
by the integration over the azimuthal phases) at largely and moderately
negative values of $x_F$, leaving active mainly the Sivers
contribution. Notice that, contrary to what happens in the $p\, p\to h
\, X$ process, no $\hat u$-channel in the partonic process is
present; moreover the variable $\hat t$ strongly depends on $\phi$,
the azimuthal phase of the Sivers effect [first term on the
r.h.s.~of Eq.~(\ref{anhn})].
\item
In Fig.~\ref{fig:an-hermes} we present our estimates, separately,
for the Sivers and Collins contributions to $A_N$ at HERMES
kinematics. More precisely, we show the Sivers effect at
$P_T=1.5$~GeV (left panel) and at $P_T=2.5$~GeV (central panel) and
the Collins effect at $P_T=2.5$~GeV (right panel). The Collins
effect at $P_T=1.5$ GeV (not shown) is almost negligible in the
kinematical region considered.
For charged pion production at $P_T$ =1.5 GeV (left panel)
the statistical uncertainty bands as resulting from our
fit~\cite{Anselmino:2008sga} are also shown.

The largest $A_N$ values obtained correspond to the $x$ region (of
the polarized proton distributions) where the Sivers functions, for
$u$ and $d$ quarks, reach their maxima. It is
interesting to note that the sizable value of $A_N$ for $\pi^-$
production (larger than the corresponding Sivers contribution to
$A_{UT}$ in SIDIS) is due to the dominance of the $d$ quark with a
small contamination from the $u$ quark.
This is related to the fact that the light-cone momentum fraction $z$
is always bigger than the maximum between $|x_F|$ and $x_T$, implying,
at moderate and large $|x_F|$, a dominance of the {\em leading}
fragmentation functions.
\begin{figure*}[t!]
\includegraphics[width=5.truecm,angle=-90]{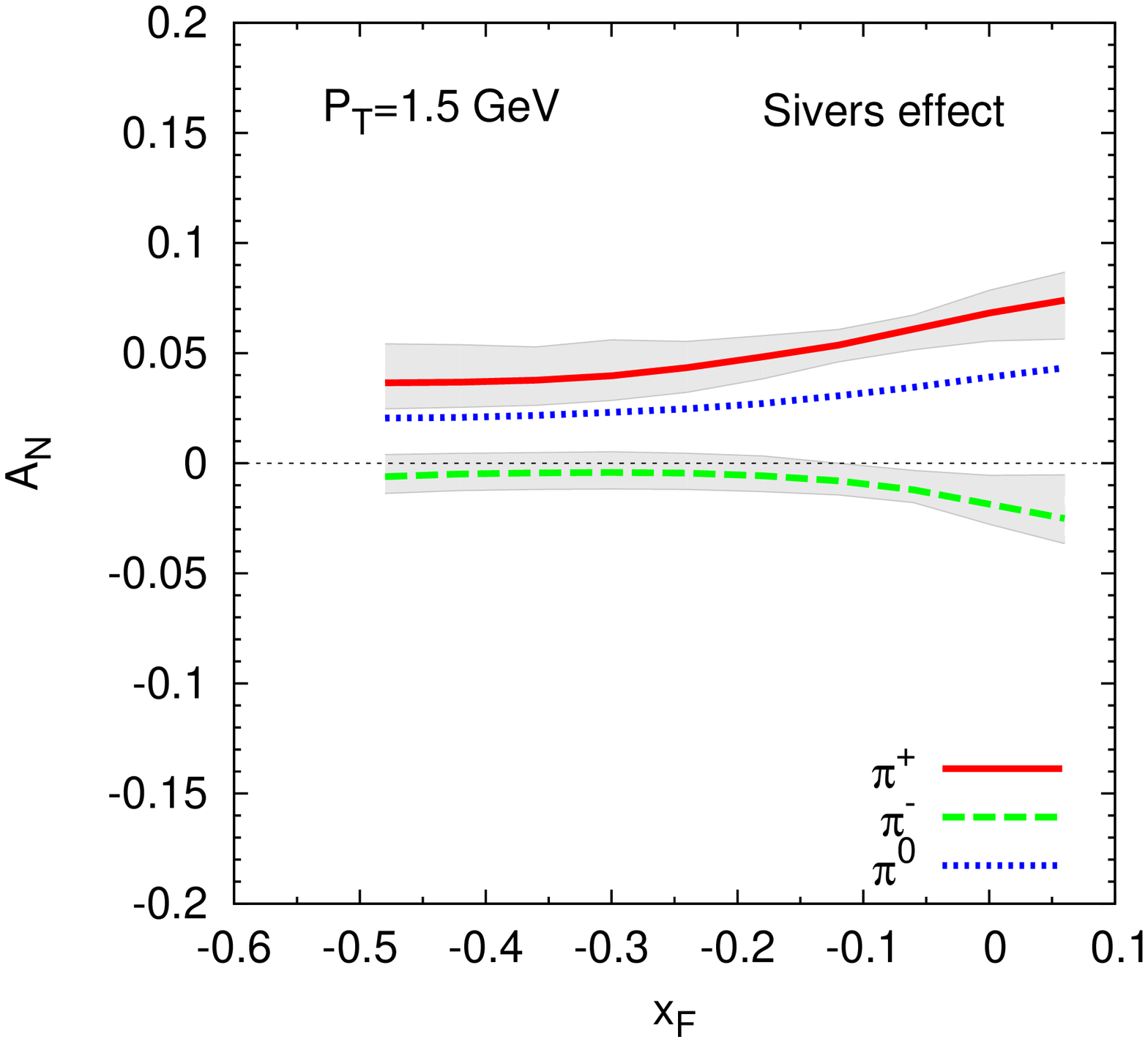}
\includegraphics[width=4.7truecm,angle=-90]{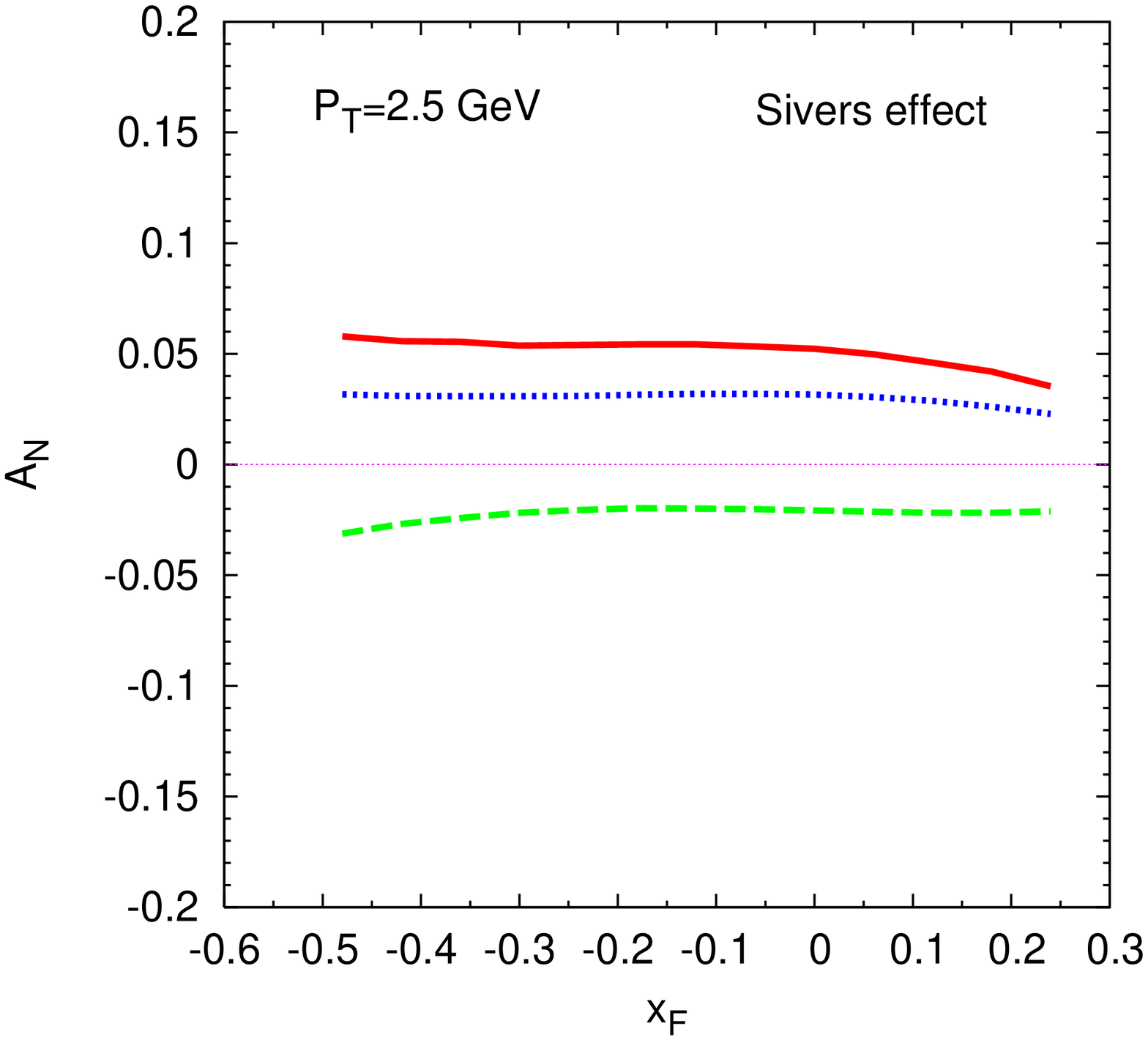}
\includegraphics[width=4.7truecm,angle=-90]{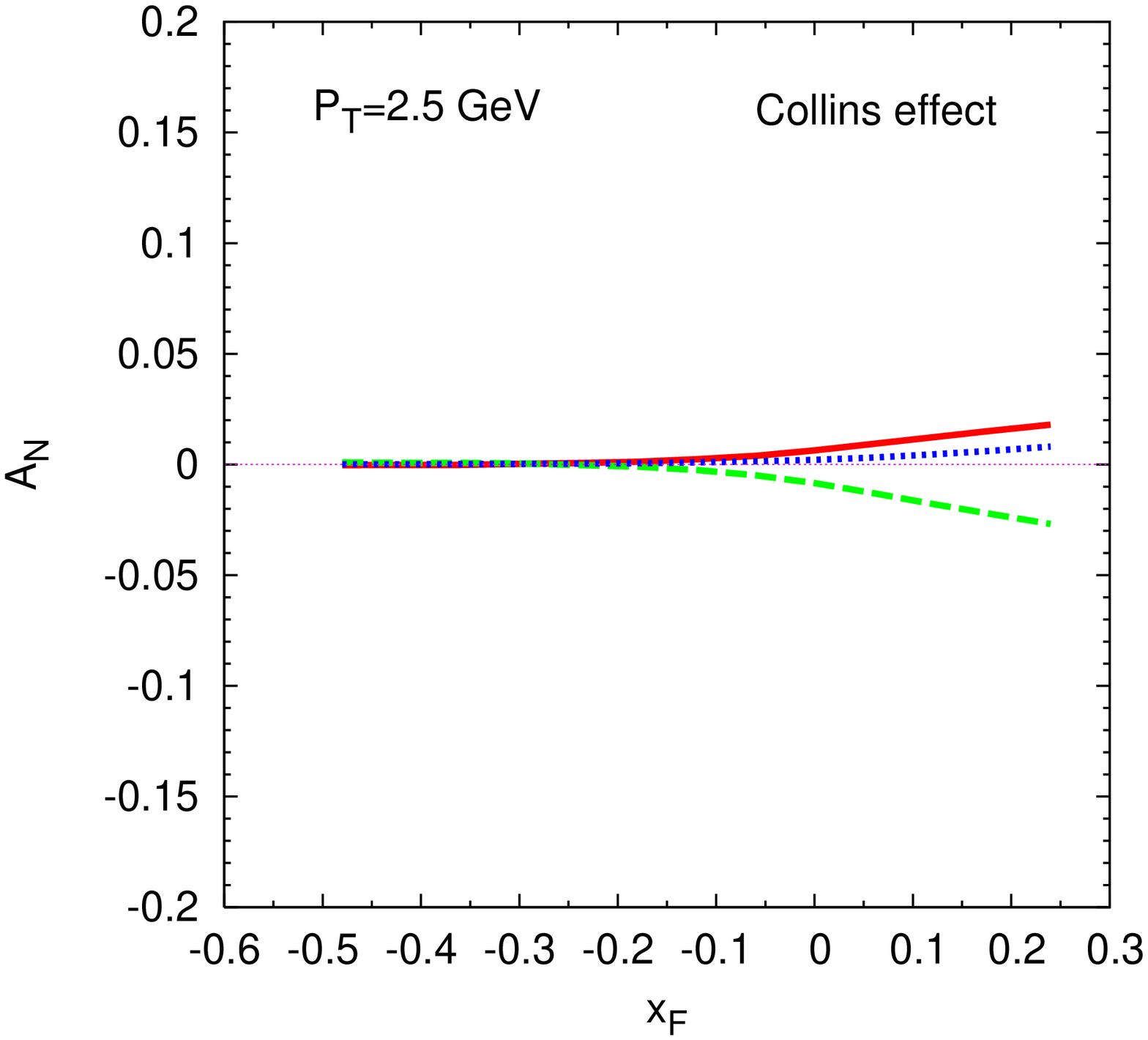}
\caption{Same as in Fig.~\ref{fig:an-hermes} but for COMPASS
kinematics ($\sqrt{s}\simeq 17$ GeV).} \label{fig:an-compass}
\end{figure*}
%
\item
In Fig.~\ref{fig:an-compass} we show the analogous results for
COMPASS kinematics. Again at $P_T$ = 1.5~GeV only the Sivers effect
gives a sizable contribution (left panel), while the Collins effect
(not shown) is compatible with zero. At $P_T$ = 2.5~GeV the Sivers
effect (central panel) dominates only in the backward region, while
in the forward region the Collins effect (right panel) becomes
sizable. For charged pion
production at $P_T$ =1.5 GeV (left panel) the statistical
uncertainty bands as resulting from our fit~\cite{Anselmino:2008sga}
are also shown. The main difference w.r.t.~$A_N$ for HERMES
kinematics at $P_T=1.5$~GeV (compare Figs.~\ref{fig:an-hermes} and
\ref{fig:an-compass}, left panels) is that at the larger COMPASS
energy ($\sqrt s \simeq 17$~GeV) the valence region for the
polarized proton, where the Sivers functions reach their maxima,
starts dominating at larger $x_F$.
\item
In Fig.~\ref{fig:an-enc} we show our results for ENC
kinematics at $\sqrt {s} = 50$~GeV. For $P_T=1.5$~GeV (left panel)
only at the upper range of the safe $x_F$ values ({\em i.e.}~$x_F\lesssim 0$)
the Sivers effect gives a sizable contribution, of the order of few
percent (the Collins effect is once again negligible). At
$P_T=2.5$~GeV both the Sivers (central panel) and the Collins (right
panel) contributions are comparable and sizable around $x_F\simeq
0.2$, therefore hardly distinguishable. This can be understood
because at such $P_T$ and energy values the valence region for the
polarized proton dominates only for $x_F>0$, where both effects are
active (see our comment on the suppression of the Collins effect for
negative $x_F$ at the beginning of this Section).
\begin{figure*}[c!h]
\includegraphics[width=5.1truecm,angle=-90]{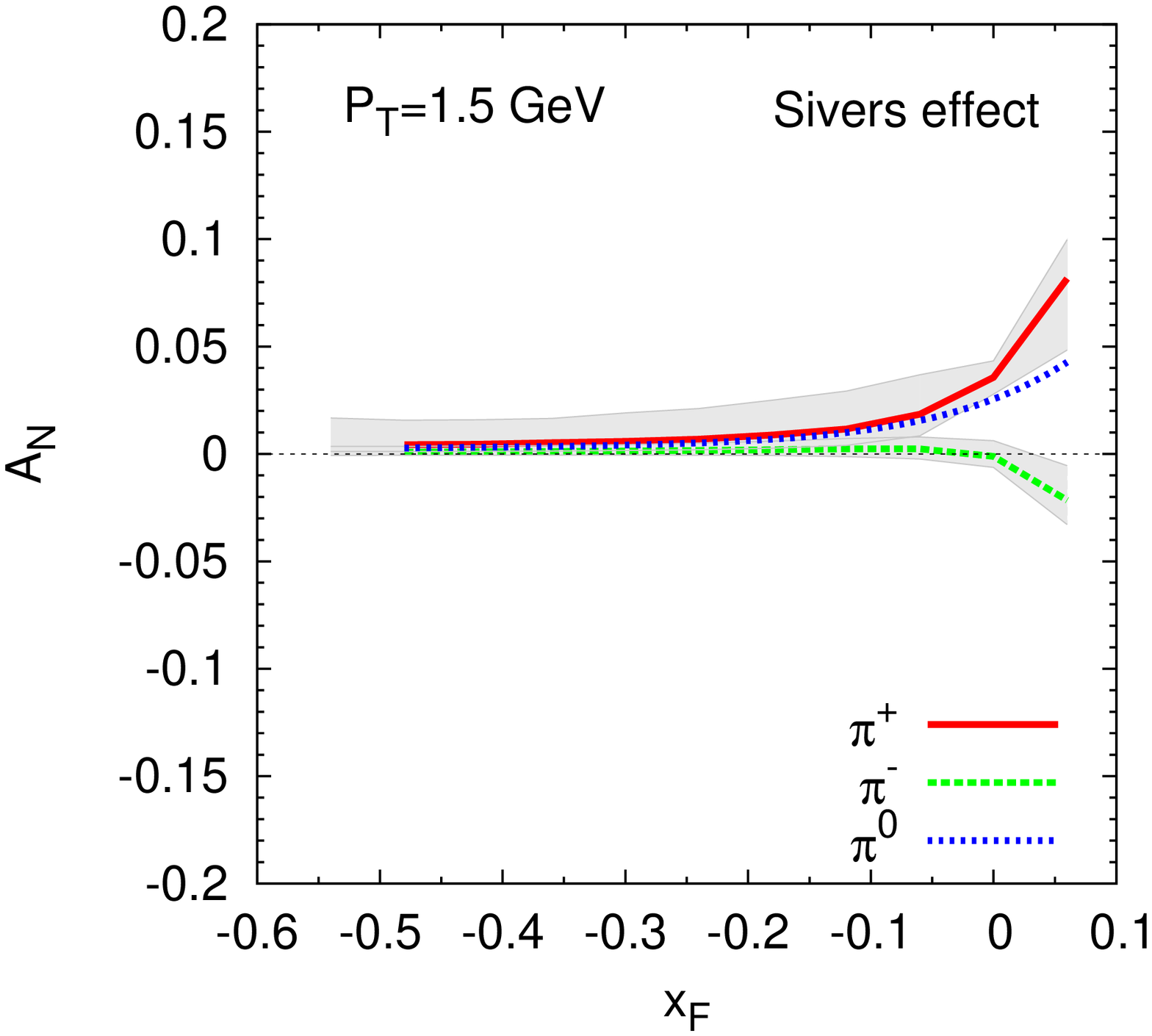}
\includegraphics[width=4.7truecm,angle=-90]{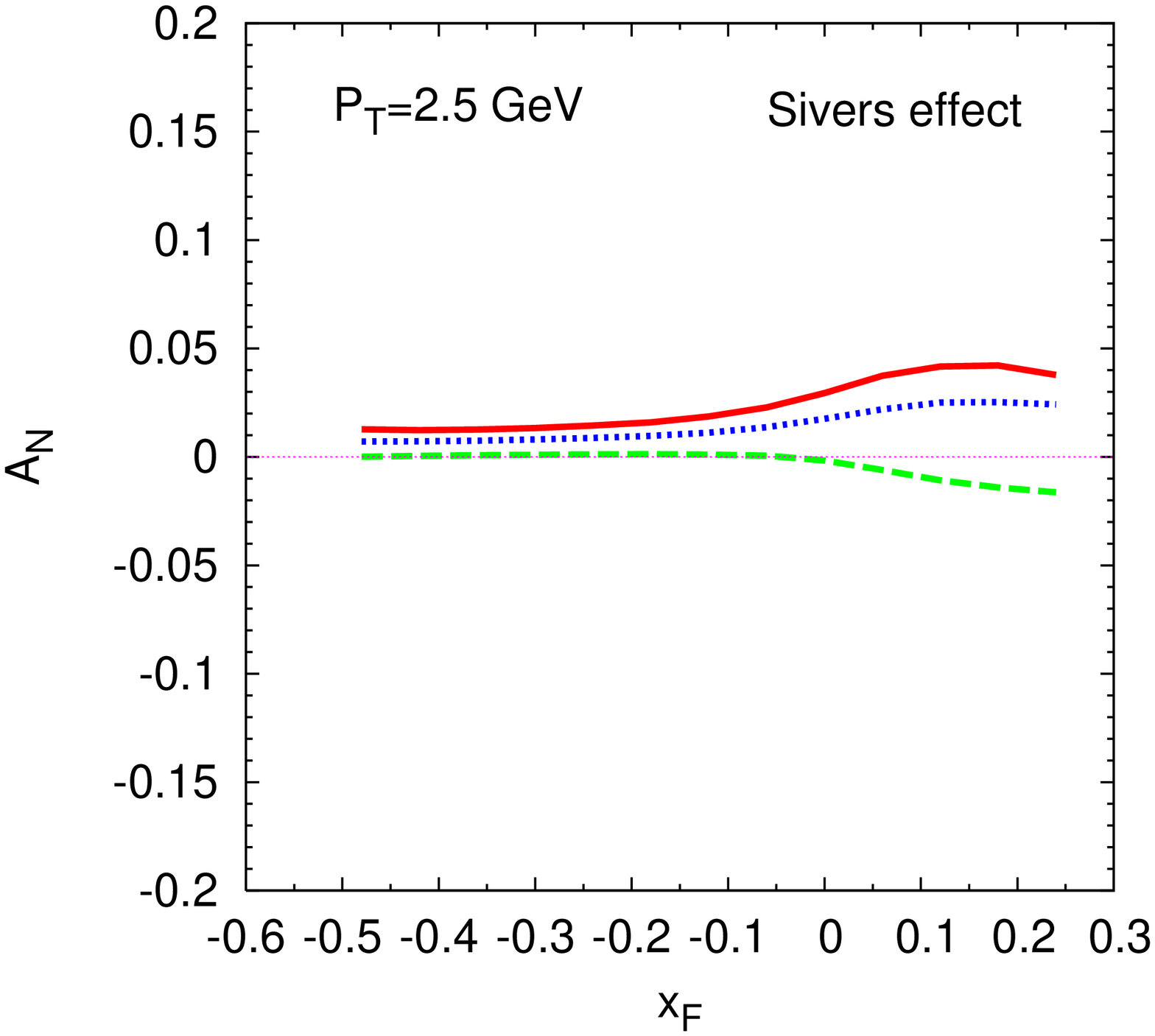}
\includegraphics[width=4.7truecm,angle=-90]{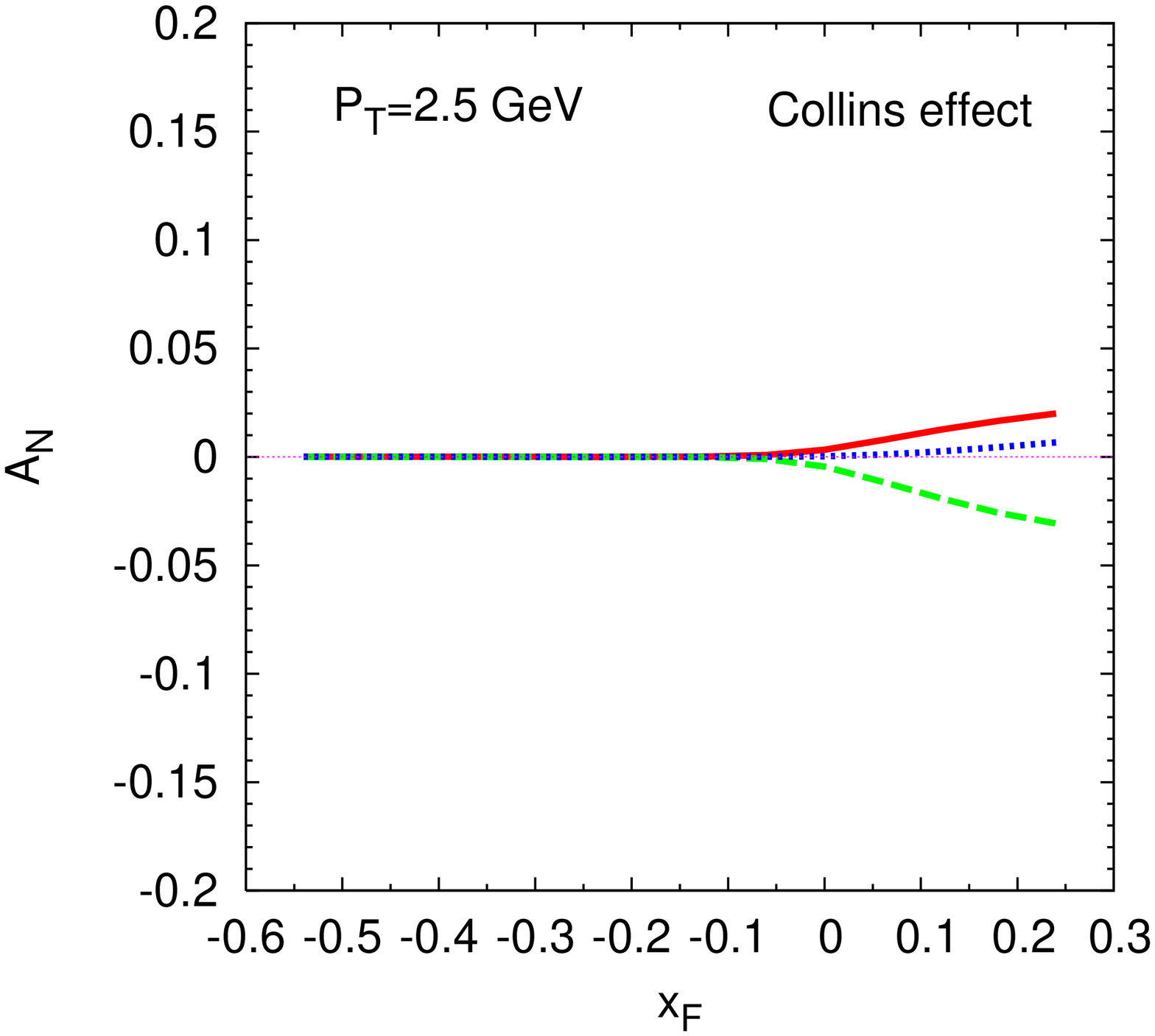}
\caption{Same as in Fig.~\ref{fig:an-hermes} but for ENC
kinematics at $\sqrt{s}= 50$~GeV.} \label{fig:an-enc}
\end{figure*}
%
\item
In Fig.~\ref{fig:an-jlab} we show analogous estimates of the Sivers
contribution to $A_N$ for JLab kinematics at the upgraded energy
$E_{\rm Lab}=12$ GeV, corresponding to a c.m.~energy $\sqrt{s} \simeq 4.9$ GeV.
Again, in order to guarantee a sufficiently large momentum transfer
we show results at $P_T=1.5$ GeV vs. $x_F \lesssim 0.1$. The results are
comparable to the corresponding estimates for HERMES kinematics, see Fig.~\ref{fig:an-compass} (left panel),
with large asymmetries (in size) for all pions.
Given the lower c.m.~energy, however, cross sections are in general smaller
than those for HERMES and COMPASS kinematics.
Larger values of $P_T$ are probably out of reach at Jlab,
while the Collins contribution is again negligible in the $x_F$ region considered.
\begin{figure*}[t!]
\includegraphics[width=5.6truecm,angle=-90]{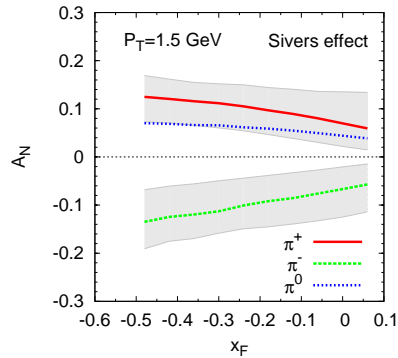}
\caption{Same as in Fig.~\ref{fig:an-hermes}, left panel, but for JLab
kinematics at $E_{\rm Lab}=12$ GeV ($\sqrt{s}\simeq 4.9$~GeV).}
\label{fig:an-jlab}
\end{figure*}
%
\begin{figure*}[t!]
\includegraphics[width=5.2truecm,angle=-90]{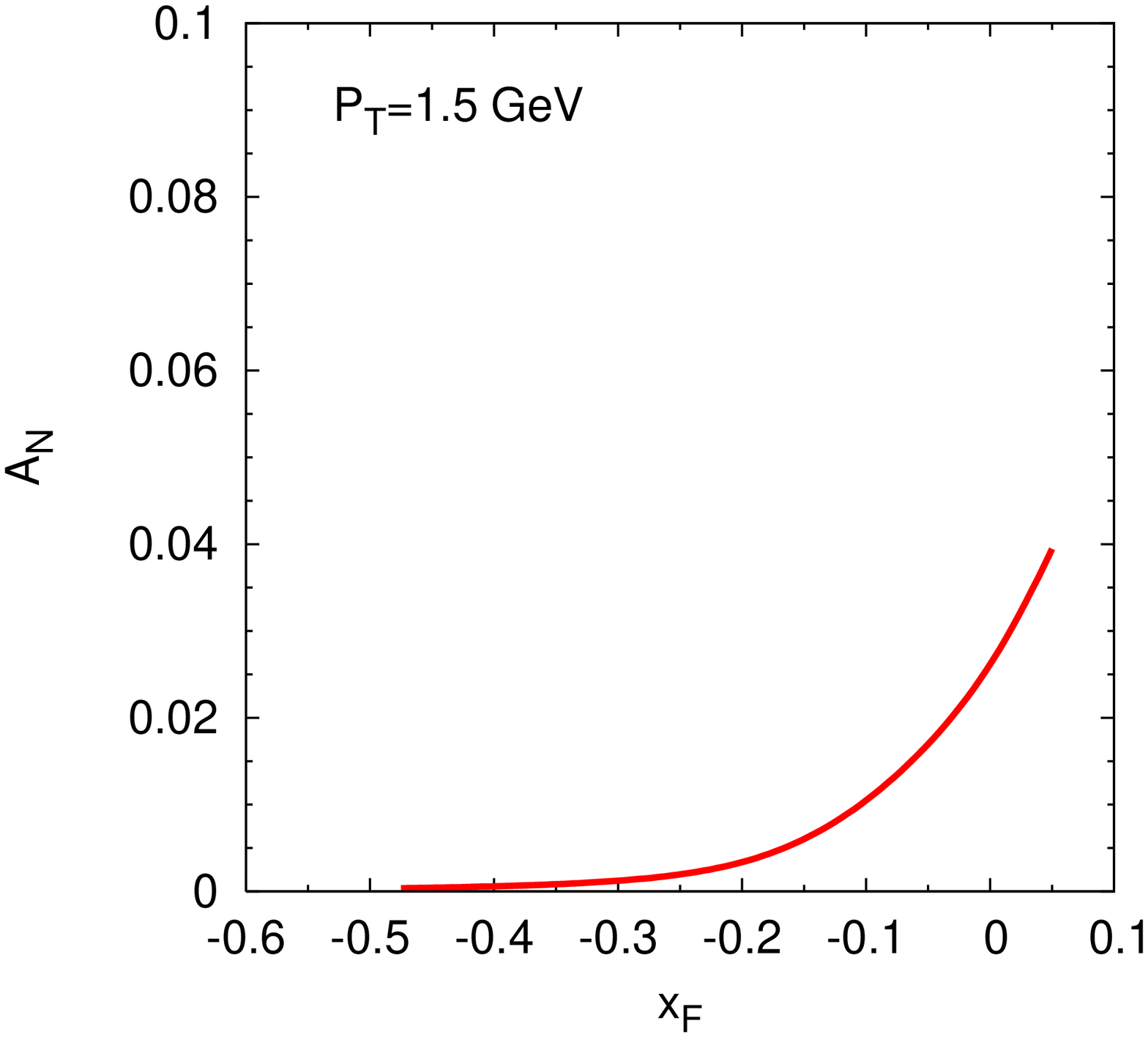}
\includegraphics[width=5.2truecm,angle=-90]{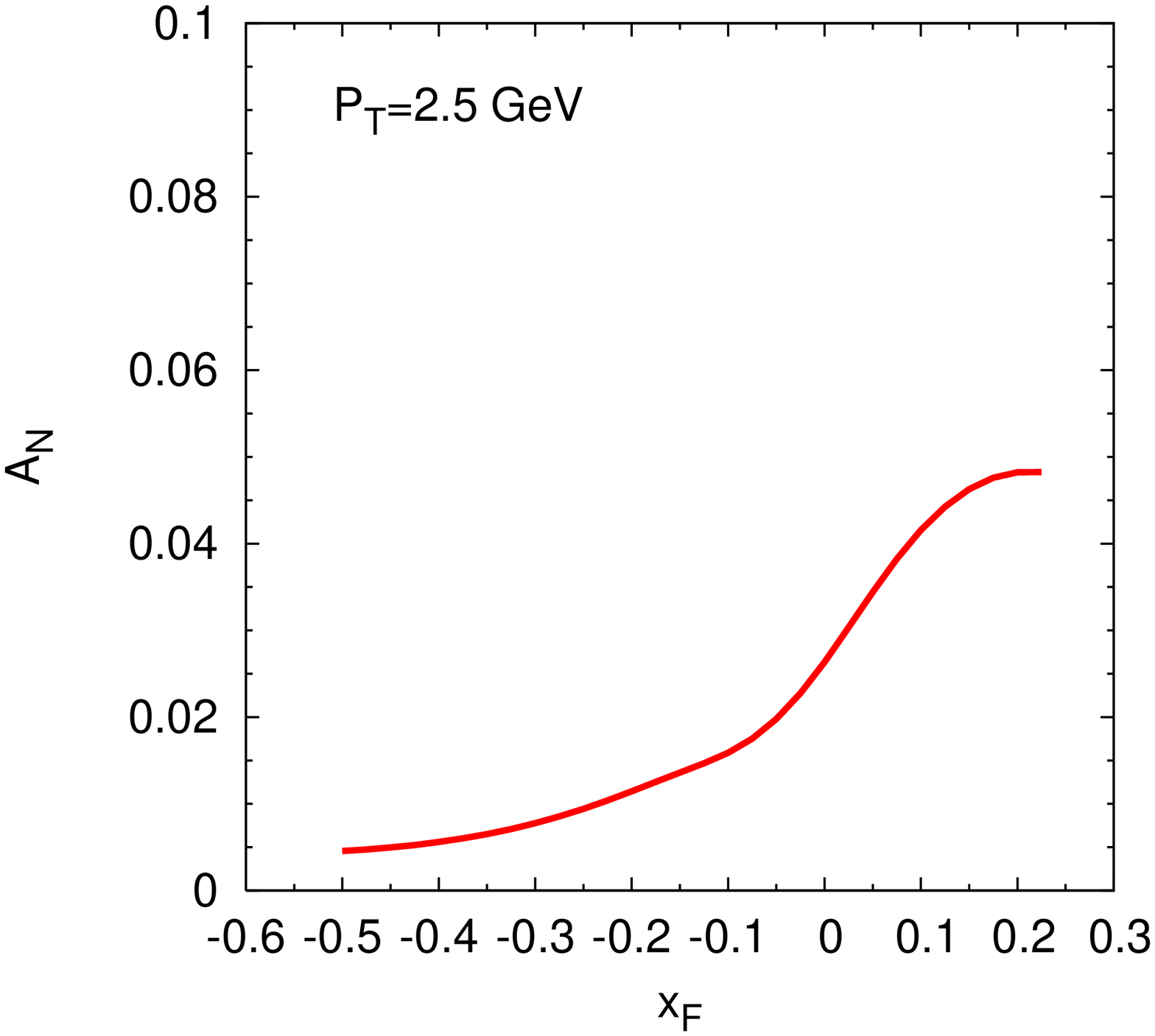}
\caption{Estimates of $A_N$ vs.~$x_F$ for the $\pup \, \ell
\to {\rm jet} + X$ process and for ENC kinematics at $\sqrt{s} = 50$ GeV.
Left panel: Sivers effect at $P_T = 1.5$~GeV;
right panel: Sivers effect at $P_T = 2.5$~GeV.
The computation has been performed according to Eqs.~(\ref{anjet}), (\ref{dsjet}) and
(\ref{ssjet}) of the text, adopting the Sivers functions of
Ref.~\cite{Anselmino:2008sga}, as extracted from SIDIS data, and
the unpolarized PDFs of Ref.~\cite{Gluck:1998xa}.}
\label{fig:jet-enc}
\end{figure*}
%
\item
Finally, in Fig.~\ref{fig:jet-enc} we show estimates of the Sivers contribution to $A_N$
for the process $p^\uparrow \ell \to {\rm jet} + X$ for ENC kinematics ($\sqrt{s}=50$ GeV) at $P_T=1.5$ GeV
(left panel) and $P_T=2.5$ GeV (right panel) vs. $x_F$.
The results are similar, both in size and shape, to the corresponding ones for neutral and positive pions,
see Fig.~\ref{fig:an-enc}, left and central panels (notice the different scale). The asymmetry is almost negligible at
negative $x_F$ and becomes sizable only at the upper range of the safe $x_F$ values.
We have found that $A_N$ becomes even smaller at larger c.m.~energies.
\end{itemize}
\section{Comments and conclusions}\label{comm}
In this paper we have presented a phenomenological study, based on the
assumption of TMD factorization, of transverse single spin asymmetries
for the inclusive production of large $P_T$ pions and jets in lepton-proton
collisions, $\pup \, \ell \to h \, ({\rm jet}) + X$. These asymmetries,
measured in the lepton-proton c.m.~frame (since the final lepton is not
observed, the $\gamma^*$--proton c.m.~frame cannot be reconstructed),
should involve the same TMD distribution and fragmentation functions
which contribute to the transverse azimuthal asymmetries
measured by the HERMES and COMPASS collaborations in the last years in
semi-inclusive deep inelastic scattering.

Using best-fit parameterizations of the TMD functions extracted from
HERMES, COMPASS and Belle data, we have shown that, in the kinematical
regions where our perturbative approach should be reliable, the asymmetries
dominantly arise from the Sivers effect in the distribution sector and
marginally from the Collins effect in the fragmentation sector
(not present in the case of jet production). We have presented results
for several kinematical configurations corresponding to present experimental
setups (HERMES and COMPASS), to the forthcoming 12~GeV upgraded JLab setup
and to a class of lepton-proton (ion) colliders
(ENC) currently under active study in the QCD
and hadron physics community. These results show that for pion
production the Sivers $A_N$ can be sizable, at least for HERMES,
COMPASS and JLab at 12~GeV kinematics. For pion and jet production at typical
energies of the proposed ENC colliders the asymmetries are much smaller
and become larger only at the boundary of the safe kinematical regions,
where, for pions, both the Sivers and the Collins contributions play
a role and the two mechanisms cannot be disentangled.

The measurement of these predicted asymmetries allows a test of the
validity of the TMD factorization, largely accepted for SIDIS processes
with two scales (small $P_T$ and large $Q$), but still much debated for
processes with only one large scale ($P_T$), like the one we are
considering here. A test of TMD factorization in such processes is of
great importance for a consistent understanding of the large SSAs
measured in the single inclusive production of large
$P_T$ hadrons in proton-proton collisions.

We stress once more that our predictions refer to large $P_T$
production, in the lepton-proton c.m.~frame, at leading
perturbative order. It implies that, in order to compare experimental
data with our results,  one has to select large $P_T$, single-jet events,
excluding those events containing a second jet in the opposite hemisphere
w.r.t.~to the primary observed jet (containing the final observed hadron).
This should avoid large $P_T$ jets (or hadrons) coming from next-to-leading
order partonic processes (hard pQCD corrections).
Although these requirements might correspond to smaller cross sections
and difficult selection procedures, we believe that the relevance of testing
TMD factorization in this simple process justifies efforts in this
direction and motivates our work.
\begin{acknowledgments}
We acknowledge support of the European Community - Research Infrastructure
Activity under the FP6 ``Structuring the European Research Area''
program (HadronPhysics, contract number RII3-CT-2004-506078).
M.A.,~M.B., and A.P.~acknowledge partial support by MIUR under Cofinanziamento
PRIN 2006.
This work is partially supported by the Helmholtz Association through
funds provided to the virtual institute ``Spin and strong QCD''(VH-VI-231).
\end{acknowledgments}
\appendix
\section{\label{kin} Kinematics}
\subsection{Hadron production}
We work in the proton-lepton center of mass frame, with the incoming
proton and lepton moving along the $Z_{\rm cm}$ axis and the outgoing
hadron emitted in the $(XZ)_{\rm cm}$ plane:
 \bea
&& p=\frac{\sqrt{s}}{2} \left(1,0,0,1 \right) \\
&& \ell= \frac{\sqrt{s}}{2} \left(1,0,0,-1 \right) \\
&& P_h = (E_h, P_T,0,P_L) \quad\quad E_h^2 = P_T^2 + P_L^2 \>,
 \eea
where $s$ is the proton-lepton c.m.~square energy and where we have
assumed all particles to be massless. The kinematical variables for
the elementary underlying process result in $(k_\perp = |\bfk_\perp|, \>
p_\perp = |\bfp_\perp| )$
\bea
&& p_q = \left(\frac{x\sqrt{s}}{2} + \frac{k^2_\perp}{2x\sqrt{s}}, \;
\bfk _\perp\;,\frac{x\sqrt{s}}{2} - \frac{k^2_\perp}{2x\sqrt{s}} \right) \\
&& \ell = \frac{\sqrt{s}}{2} \left(1,0,0,-1 \right) \\
&& p^\prime_q = \frac{E_h+\sqrt{E_h^2-p_\perp^2}} {2z}
\Big[1,\frac{1}{\sqrt{E_h^2-p_\perp^2}} (P_T-p_\perp^x,-p_\perp^y,P_L-p_\perp^z )
\Big]\\
&& \ell' = p_q + \ell - p'_q \>,
\eea
with $\bfk_\perp$ being the intrinsic transverse momentum of parton $q$ inside
the parent proton and $\bfp_\perp$ being the intrinsic transverse momentum of
the detected final hadron $h$ with respect to the fragmenting parton $q^\prime$.
The expression for $p^\prime_q$ has been obtained by requiring $z$ to be the
light-cone momentum fraction of the emitted hadron, $z=\tilde{P}_h^+/\tilde{p}_q^{\prime +}$
as defined in the helicity frame of the fragmenting quark $q^\prime$, which we
will denote as $\tilde{S}$.
With this kinematics, the partonic Mandelstam invariants are
\bea
\hat{s} &=& xs \nonumber \\
\hat{t} &=& -\frac{x\sqrt{s}}{2z}\left( 1+ \frac{E_h}
{\sqrt{E_h^2-p_\perp^2}}\right)
\left[ \left( 1 + \frac{\kt^2}{x^2s}\right)\sqrt{E_h^2 - p_\perp^2} -
\left( 1-\frac{\kt^2}{x^2s}\right) (P_L-p_\perp^z) \right. \nonumber \\
&& \hskip 110 pt
\left. - \, \frac{2\kt^x \, (P_T - p_\perp^x) - 2\kt^y \, p_\perp^y}
{x\sqrt s} \right]\nonumber \\
\hat{u} &=& -\frac{\sqrt s}{2z} \left( 1 + \frac{E_h}
{\sqrt{E_h^2 - p_\perp^2}}\right) \left(\sqrt{E_h^2 - p_\perp^2}
+ P_L - p_\perp^z \right)\>.
\eea

Notice that the orthogonality between $\bfp_q^\prime$ and
$\bfp_\perp$, explicitly guaranteed through the delta function
$\delta(\bfp_{\perp} \cdot \hat{\bfp}'_q)$ in Eq.~(\ref{anh}),
allows us to fix one component of the vector $\bfp_\perp$ in terms
of all the others; in particular it gives
\be
|\bfp_\perp|^2 = P_T\,p_\perp^x + P_L\,p_\perp^z \;\;\;\;
\Longrightarrow \;\; \;\;
p_\perp^y = \pm \sqrt{P_T\,p_\perp^x + P_L\,p_\perp^z -(p_\perp^x)^2 -
(p_\perp^z)^2 } \,. \label{p-perp-y}
\ee
Similarly, the other delta function in Eq.~(\ref{anh}),
$\delta(\hat{s} +\hat{t} + \hat{u})$, can be used to perform the integration
over the light-cone fraction $z$ fixing
\bea
z = \frac{1}{2x \sqrt s} \left( 1 + \frac{E_h}{\sqrt{E_h^2-p_\perp^2}}\right)
&& \!\!\!\!\! \left[ \left( 1+x+\frac{\kt^2}{x s}\right)
\sqrt{E_h^2-p_\perp^2} + \left( 1-x+\frac{\kt^2}{x s}\right)
(P_L-p_\perp^z) \nonumber \right. \\
&&\!\!\!\!\!\!\!\! \left. - \frac{2\kt^x(P_T-p_\perp^x) - 2\kt^y
\,p_\perp^y }{\sqrt s} \right], \eea
with $p_\perp^y$ given by Eq.~(\ref{p-perp-y}).

The angle $\phi_q^h$, which identifies the direction of $\bfp_\perp$ around
$\bfp^\prime_q$, can be expressed in terms of the $\bfp_\perp$ components,
$p_\perp^x$, $p_\perp^y$ and $p_\perp^z$, simply by noticing that in the
helicity frame of parton $q^\prime$ (where the $\tilde{Z}$ axis coincides with the
direction of $\bfp_q^\prime$) this angle is the azimuth of $\bfp_\perp$,
that is:
\be
\sin\phi_q^h = \hat{\bfp}_\perp \cdot \tilde{\bm{Y}}
\quad\quad\quad\quad
\cos\phi_q^h = \hat{\bfp}_\perp \cdot \tilde{\bm{X}} \>.
\label{phiqh}
\ee
The helicity frame $\tilde{S}$ of parton $q^\prime$ can be reached by
performing two rotations, as explained in Appendix C of
Ref.~\cite{Anselmino:2005sh}, in the following way
\bea
&&\tilde{\bm{Z}}=\hat{\bfp}^\prime_q = \frac{1}{\sqrt{E_h^2-p_\perp^2}}
(P_T-p_\perp^x,-p_\perp^y,P_L-p_\perp^z ) \label{tilde-axes-X} \\
&&\tilde{\bm{Y}}=\hat{\bm{Z}}_{\rm cm} \times \hat{\bfp}_{qT}^\prime =
\frac{(p_\perp^y,P_T-p_\perp^x,0)}{\sqrt{E_h^2-p_\perp^2 -
(P_L - p_\perp^z)^2}} \label{tilde-axes-Y} \\
&&\tilde{\bm{X}} = \tilde{\bm{Y}} \times \tilde{\bm{Z}} =
\frac{[(P_T-p_\perp^x)(P_L-p_\perp^z), \>
- p_\perp^y (P_L-p_\perp^z), \>
- (p_\perp^y)^2 - (P_T-p_\perp^x)^2]}
{\sqrt{E_h^2 - p_\perp^2 - (P_L - p_\perp^z)^2} \>
\sqrt{E_h^2 - p_\perp^2}} \>, \label{tilde-axes-Z}
\eea
where $\hat{\bfp}^\prime_{qT}$ is given by the transverse components
of $\bfp^\prime_q$ in the center of mass reference frame, $S$:
\be
\hat{\bfp}^\prime_{qT} = \frac{1}{\sqrt{E_h^2-p_\perp^2 -
(P_L - p_\perp^z)^2}}(P_T-p_\perp^x,-p_\perp^y,0),
\ee
and $p_\perp^y$ is fixed by the orthogonality condition of
Eq.~(\ref{p-perp-y}).

By replacing Eqs.~(\ref{tilde-axes-X})-(\ref{tilde-axes-Z}) into
Eq.~(\ref{phiqh}) we find
\bea
&&\sin\phi_q^h = \frac{p_\perp^y}{p_\perp}\,\frac{P_T}{\sqrt{E_h^2-p_\perp^2 -
(P_L - p_\perp^z)^2}} \nonumber \\
&&\cos\phi_q^h = -\frac{p_\perp^z}{p_\perp}\,\frac{\sqrt{E_h^2-p_\perp^2}}
{\sqrt{E_h^2-p_\perp^2 - (P_L - p_\perp^z)^2}}.
\eea
Alternatively, in terms of angles instead of components we can write
\bea
&&\sin\phi_q^h = -\frac{P_T}{p_\perp}\,\sin\phi^\prime \nonumber \\
&&\cos\phi_q^h = -\frac{p_\perp^z}{p_\perp}\frac{1}{\sin\theta^\prime} =
-\frac{\cos\theta_\perp}{\sin\theta^\prime} \>,
\eea
where $\phi', \theta'$ are the azimuthal, polar angles of $\bfp_q'$
and $\theta_\perp$ is the polar angle of $\bfp_\perp$ in our c.m.~reference
frame.

{F}inally, the $\cos(\phi^\prime+\phi_q^h)$ azimuthal dependence of the Collins
effect, see Eq.~(\ref{anhn}), can be expressed as
\be
\cos(\phi^\prime + \phi_q^h) = \frac{p_\perp^z\,(p_\perp^x - P_T)
\sqrt{E_h^2-p_\perp^2} + (p_\perp^y)^2 P_T}
{p_\perp \, [E_h^2-p_\perp^2 - (P_L - p_\perp^z)^2]} \,,
\ee
or, more simply, in terms of angles
\be
\cos(\phi^\prime+\phi_q^h) = \frac{P_T}{p_\perp}\,\sin ^2\phi^\prime -
\cos\theta_\perp \frac{\cos\phi^\prime}{\sin\theta^\prime}\> \cdot
\ee
\subsection{\label{kin-jet} Jet production}
The 4-momenta involved, in our reference frame and neglecting all masses, are
\bea
&& p= \frac{\sqrt{s}}{2} \left(1,0,0,1 \right) \quad\quad
\ell=\frac{\sqrt{s}}{2} \left(1,0,0,-1 \right) \\
&& p_q = \left(\frac{x\sqrt{s}}{2} + \frac{k^2_\perp}{2x\sqrt{s}},
\;\bfk _\perp\;,\frac{x\sqrt{s}}{2} - \frac{k^2_\perp}{2x\sqrt{s}} \right) \\
&& p'_q = P_j = (E_j, P_T,0,P_L) \quad\quad E_j^2 = P_T^2 + P_L^2 \>,
\eea
so that the partonic Mandelstam invariants are given by
\bea
&&\hat s = xs\\
&&\hat t = 2P_T k_\perp^x - x\sqrt{s} \left[E_j - P_L +
\frac{k_\perp^2}{x^2s} (E_j + P_L)\right]\\
&&\hat u = -\sqrt{s}\,(E_j + P_L) \>. \eea
Notice that there is no linear $k_\perp^y$ dependence in these
variables and, as a consequence, in the elementary amplitudes $\hat
M^0_{1,2}$. The delta function ensuring $\hat s+\hat t+\hat u = 0$
can be used to perform the integration on $k_\perp^x$ in
Eq.~(\ref{anjet}):
\be
k_\perp^x = x \sqrt s \left[ \frac{P_T}{E_j + P_L} \pm
\sqrt{\frac{\sqrt s}{E_j + P_L} - \frac{1}{x} - \frac{(k_\perp^y)^2}{x^2 s}}
\> \right] \> , \label{kx}
\ee
which implies
\be
x_{\rm min} = \frac{E_j + P_L}{2\sqrt{s}} \left[ 1 +
 \sqrt{1+\frac{4(k_\perp ^y)^2}{\sqrt{s}(E_j+P_L)}} \> \right]\,.
\label{xmin}
\ee
Note that the term proportional to $\cos\phi_S$ in
Eq.~(\ref{defsivnoi2}), being odd in $k_\perp^y$, vanishes when
integrating over $k_\perp^y$, resulting, as it should, in a SSA
proportional to $\sin\phi_S$.
\section{\label{spinors} Spinors and helicity amplitudes}
We compute the helicity amplitudes for the non-planar $q \, \ell \to q \, \ell$
process exploiting the well known spinor helicity technique (see,
for example, Refs.~\cite{Dixon:1996wi,Leader:2001gr}). To be
precise, we adopt the phase convention and gamma matrix
representation of Ref.~\cite{Leader:2001gr}; that is, our helicity
spinors for a massless Dirac particle with 4-momentum $k = (k^0,
k^x, k^y, k^z)$ and helicity $\pm 1/2$ are given by:
\be \label{eq:explicitspinor_weil}
u_+(k) = v_-(k) =
  \left( \matrix{ \sqrt{k^+} e^{-i\phi/2}\cr
    \sqrt{k^-} e^{i\phi/2} \cr
   0 \cr
   0 \cr} \right) , \hskip3mm
u_-(k) = v_+(k) =
  \left( \matrix{ 0 \cr
                  0 \cr
                 - \sqrt{k^-} e^{-i\phi/2}\cr
    \sqrt{k^+} e^{i\phi/2}\cr} \right) ,
\ee
where
\be \label{eq:phasekdef}
e^{\pm i\phi}\ \equiv\
  { k^x \pm ik^y \over \sqrt{(k^x)^2+(k^y)^2} }
\ =\  { k^x \pm ik^y \over \sqrt{k^+k^-} }\ ,
\qquad k^\pm\ =\ k^0 \pm k^z.
\ee
The two independent helicity amplitudes
$\hat M_{\la_3,\la_4; \la_1, \la_2}$ for the
$q(k_1,\la_1) + \ell(k_2, \la_2) \to q(k_3, \la_3) + \ell(k_4, \la_4)$
elementary lowest order QED interaction are given by:
\bea
\hat M_{++;++} = \hat M^*_{--;--} &=&
- \, 2 \, \frac{e_q e^2}{\hat t} \spb{3}.{4} \spa{1}.{2} \\
\hat M_{+-;+-} = \hat M^*_{-+;-+} &=&
+ \, 2  \, \frac{e_q e^2}{\hat t} \spb{2}.{3} \spa{1}.{4} \>,
\eea
with
\be
\bar u_-(k_i)\,u_+(k_j)\equiv
\spa{i}.{j} = - \spb{i}.{j}^* =
\sqrt{k_i^+ k_j^-} \, e^{- i(\phi_{i} -\phi_{j})/2 } -
\sqrt{k_i^- k_j^+} \, e^{i(\phi_{i} - \phi_{j})/2} \>.
\label{<ij>}
\ee
Eq.~(\ref{<ij>}) can be rewritten as~\cite{Dixon:1996wi}:
\be
\spa{i}.{j} = -e^{- i(\phi_{i} + \phi_{j})/2} \left[
\sqrt{k_i^- k_j^+} \, e^{i \phi_{i}} -
\sqrt{k_i^+ k_j^-} \, e^{i \phi_{j}} \right]
= -e^{- i(\phi_{i} + \phi_{j})/2} \,
\sqrt{|s_{ij}|} \, e^{i\phi_{ij}},
\label{eq:posenergyexplicit}
\ee
where  $s_{ij}\ =\ (k_i+k_j)^2\ =\ 2 k_i\cdot k_j$, and
\be \label{eq:phiijdef}
\cos\phi_{ij}\ =\ { k_i^x k_j^+ - k_j^x k_i^+
             \over \sqrt{|s_{ij}| \, k_i^+ k_j^+} }
\ , \qquad
\sin\phi_{ij}\ =\ { k_i^y k_j^+ - k_j^y k_i^+
             \over \sqrt{|s_{ij}| \, k_i^+ k_j^+} }
\ , \qquad \phi_{ij} =  \phi_{ji} + \pi \>.
\ee

With our kinematical configuration ($\phi_2 = \pi$) we obtain:
\bea
\hat M_{++;++} = \hat M^*_{--;--} &=& - 8 \, \pi \, e_q \, \alpha \,
\frac{\hat s}{\hat t} \> e^{-i\phi_{34}} \,
e^{i(\phi_{3} +\phi_{4} - \phi_{1} + \pi)/2}
\label{m++} \\
\hat M_{+-;+-} = \hat M^*_{-+;-+} &=& 8 \, \pi \, e_q \, \alpha \,
\frac{\hat u}{\hat t} \> e^{i \phi_{14}} \,
e^{- i(\phi_{1} +\phi_{4} - \phi_{3} + \pi)/2} \>.
\label{m+-}
\eea
In addition, one can show that $\phi_{34} = \phi_{14}$.

Notice that the combinations of helicity amplitudes contributing
to the SSA, Eq.~(\ref{ds1}), are simply given by:
\bea
|\hat M_{++;++}|^2 &\equiv& |\hat M_1^0|^2 = 64 \, \pi^2 \alpha^2 e_q^2 \>
\frac{\hat s^2}{\hat t^2} \label{m1m1} \\
|\hat M_{+-;+-}|^2 &\equiv& |\hat M_2^0|^2 = 64 \, \pi^2 \alpha^2 e_q^2 \>
\frac{\hat u^2}{\hat t^2} \label{m2m2} \\
\hat M_{++;++} \, \hat M_{-+;-+}^* &=& 64 \, \pi^2 \alpha^2 e_q^2 \>
\frac{\hat s(-\hat u)}{\hat t^2} \> e^{-i(\phi_1 - \phi_3)} \>. \label{m1m2}
\eea

In the (transversity) $\otimes$ (Collins) contribution to the SSA,
the phase dependence of the last term above ($\phi_1 - \phi_3 = \phi
- \phi'$) combines with the $\bfk_\perp$ phase in the transversity
distribution ($\phi$) and the Collins function phase ($\phi_q^h$),
resulting in the simple expression given in Eq.~(\ref{ds1}).
\section{\label{aut} Details for the computation of $A_{TU}^{\sin\phi_S}$}
In this Section we show some details of the explicit calculation of the
transverse single spin asymmetry $A_{TU}^{\sin\phi_S}$, Eq.~(\ref{ATU}),
for the process $p^\uparrow\, \ell \to h \,X$, starting from the general expression
for the polarized cross section given in Eq.~(\ref{gensig}). By performing the
sum over all the helicity indices and taking into account that the helicity
density matrix of a quark $q$ can be written in terms of the quark
polarization vector components, $\bfP ^q = (P ^q_x, P ^q_y,P ^q_z)$, as
\be
\rho_{\la^{\,}_q, \la^{\prime}_q}^{q/p,S} =
{\left(
\begin{array}{cc}
\rho_{++}^{q} & \rho_{+-}^{q} \\
\rho_{-+}^{q} & \rho_{--}^{q}
\end{array}
\right)}_{\!\! p,S} = \>
\frac{1}{2}\,{\left(
\begin{array}{cc}
1+P^q_z & P^q_x - i P^q_y \\
 P^q_x + i P^q_y & 1-P^q_z
\end{array}
\right)}_{\!\!p,S},
\label{rho-q}
\ee
one obtains, for a spinless hadron $h$,
\bea
\frac{E_h \, {\rm d}\sigma^{(p,S) + \ell \to h + X}}
{{\rm d}^{3} \bm{P}_h} &=& \sum_{q}
\int \frac{{\rm d}x \, {\rm d}z}{16 \, \pi^2 x \, z^2  s} \;
{\rm d}^2 \bfk_{\perp} \, {\rm d}^3 \bfp_{\perp}\,
\delta(\bfp_{\perp} \cdot \hat{\bfp}'_q) \, J(p_\perp)
\> \delta(\hat s + \hat t + \hat u) \label{gensig1} \\
&& \hskip -80pt  \times \, \frac{1}{2} \, \left\lbrace\,
\hat f_{q/p,S}(x,\bfk_{\perp}) \> (|\hat M_{++;++}|^2 +|\hat M_{-+;-+}|^2 )
D_{h/q}(z,p_{\perp})  \right.\nonumber \\
&& \hskip -82pt \left. +
\left[ \, P_y^q \, \hat f_{q/p,S}(x,\bfk_{\perp}) \>
[{\rm Re}(\hat M_{++;++}\hat M^*_{-+;-+}) \cos\phi_q^h -
 {\rm Im}(\hat M_{++;++}\hat M^*_{-+;-+}) \sin\phi_q^h ]
\right. \right. \nonumber  \\
&& \hskip -80pt \left.\left. \> \> \> - \> \> P_x^q \, \hat
f_{q/p,S}(x,\bfk_{\perp}) \> [{\rm Im}(\hat M_{++;++}\hat
M^*_{-+;-+}) \cos\phi_q^h +
 {\rm Re}(\hat M_{++;++}\hat M^*_{-+;-+}) \sin\phi_q^h ]\, \right]
 \Delta^N\! D_{h/\qup}(z,p_{\perp})\,
\right\rbrace \>. \nonumber
\eea

In the above expression we have already extracted from the fragmentation
functions $\hat{D}_{\lambda^{}_{q},\lambda^{\prime}_{q}}(z,\bfp_\perp)$
their azimuthal dependence and exploited their parity properties
(see Ref.~\cite{Anselmino:2005sh} for details):
\bea
&& \hat D_{++}(z,\bfp_{\perp}) = \hat D_{--}(z,\bfp_{\perp}) =
D_{h/q}(z,p_{\perp}) \\
&& \hat D_{+-}(z,\bfp_{\perp}) = D_{+-}(z,p_{\perp}) \, e^{i\phi_q^h} =
\frac{i}{2}\Delta^N\! D_{h/\qup}(z,p_{\perp})\, e^{i\phi_q^h} \\
&& \hat D_{+-}(z,\bfp_{\perp}) = [\hat D_{-+}(z,\bfp_{\perp})]^* \>.
\eea

When computing the azimuthal asymmetry one has the difference of cross sections
with opposite transverse spin, ${\rm d}\sigma(\phi_S) - {\rm d}\sigma(\phi_S+\pi)$;
using Eq.~(\ref{defsivnoi}) and the definitions~\cite{Anselmino:2005sh}
\bea
&& P_y^q \hat f_{q/p,{\bfS}_T}(x,\bfk_{\perp}) -
P_y^q \hat f_{q/p,-{\bfS}_T}(x,\bfk_{\perp}) =
\Delta \hat f_{s_y/{\bfS}_T}(x,\bfk_{\perp}) -
\Delta \hat f_{s_y/-{\bfS}_T}(x,\bfk_{\perp}) =
2 \Delta^- \hat f_{s_y/{\bfS}_T}(x,\bfk_{\perp}) \nonumber \\
&& P_x^q \hat f_{q/p,{\bfS}_T}(x,\bfk_{\perp}) -
P_x^q \hat f_{q/p,-{\bfS}_T}(x,\bfk_{\perp}) =
\Delta \hat f_{s_x/{\bfS}_T}(x,\bfk_{\perp}) -
\Delta \hat f_{s_x/-{\bfS}_T}(x,\bfk_{\perp}) =
2 \Delta \hat f_{s_x/{\bfS}_T}(x,\bfk_{\perp}) \>,
\nonumber \\
\eea
one obtains
\bea
{\rm d}\sigma(\phi_S) - {\rm d}\sigma(\phi_S+\pi) &=& \sum_{q}
\int \frac{{\rm d}x \, {\rm d}z}{16 \, \pi^2 x \, z^2  s} \;
{\rm d}^2 \bfk_{\perp} \, {\rm d}^3 \bfp_{\perp}\,
\delta(\bfp_{\perp} \cdot \hat{\bfp}'_q) \, J(p_\perp)
\> \delta(\hat s + \hat t + \hat u) \nonumber  \\
&& \hskip -105pt \times \, \biggl\{\, \frac{1}{2} \,
\Delta \hat f_{q/p,{\bfS}_T}(x,\bfk_{\perp}) \>
(|\hat M_{++;++}|^2 +|\hat M_{-+;-+}|^2 ) \,
D_{h/q}(z,p_{\perp}) \nonumber \\
&& \hskip -105pt  +  \,
\left[ \, \Delta^- \hat f_{s_y/{\bfS}_T}(x,\bfk_{\perp})
[{\rm Re}(\hat M_{++;++} \hat M^*_{-+;-+}) \, \cos\phi_q^h -
 {\rm Im}(\hat M_{++;++} \hat M^*_{-+;-+}) \,  \sin\phi_q^h ]
\right. \nonumber \\
&& \hskip -105pt \left. \>\>\>+ \>\> \Delta \hat
f_{s_x/{\bfS}_T}(x,\bfk_{\perp}) [{\rm Im}(\hat M_{++;++} \hat
M^*_{-+;-+}) \, \cos\phi_q^h +
 {\rm Re}(\hat M_{++;++} \hat M^*_{-+;-+}) \, \sin\phi_q^h ]\,
\right] \Delta^N\! D_{h/\qup}(z,p_{\perp}) \, \biggr\} \>.
 \label{num1} \eea

{F}inally, using Eqs.~(\ref{ampl1}), (\ref{ampl2}), (\ref{defsivnoi}),
(\ref{defsivnoi2}), (\ref{m1m1})-(\ref{m1m2}) and the
relations~\cite{Anselmino:2005sh}
\bea \Delta^- \hat f_{s_y/{\bf S}_T}(x,\bfk_{\perp}) &=& \left[
h_1(x,k_{\perp})-\frac{k_{\perp}^2}
{2M^2} \, h_{1T}^\perp(x,k_{\perp})\right] \sin(\phi_S - \phi) \\
\Delta \hat f_{s_x/{\bf S}_T}(x,\bfk_{\perp}) &=& \left[
h_1(x,k_{\perp})+\frac{k_{\perp}^2} {2M^2} \,
h_{1T}^\perp(x,k_{\perp})\right] \cos(\phi_S - \phi) \>, \eea
yields:
\bea
{\rm d}\sigma(\phi_S) - {\rm d}\sigma(\phi_S+\pi) &=& \sum_{q}
\int \frac{{\rm d}x \, {\rm d}z}{16 \, \pi^2 x \, z^2  s} \;
{\rm d}^2 \bfk_{\perp} \, {\rm d}^3 \bfp_{\perp}\,
\delta(\bfp_{\perp} \cdot \hat{\bfp}'_q) \, J(p_\perp)
\> \delta(\hat s + \hat t + \hat u)  \nonumber \\
&\times& \,\left\lbrace \, \frac{1}{2} \, \Delta^N\!
f_{q/\pup}(x,k_{\perp}) \, \sin(\phi_S-\phi) \> ( |\hat M_1^0|^2 +
|\hat M_2^0|^2) \>  D_{h/q}(z,p_{\perp})
\right. \nonumber\\
&&\left. +  \>\> h_{1q}(x,k_{\perp}) \> \sin(\phi_S - \phi^\prime -
\phi_q^h) \> \hat M_1^0 \hat M_2^0 \> \Delta^N\!
D_{h/\qup}(z,p_{\perp})
\right. \nonumber \\
&& \left. - \>\> \frac{k_{\perp}^2}{2M^2}\,h_{1T}^{\perp
q}(x,k_{\perp}) \sin(\phi_S - 2\phi + \phi^\prime + \phi_q^h ) \>
\hat M_1^0 \hat M_2^0 \> \Delta^N\! D_{h/\qup}(z,p_{\perp}) \,
\right\rbrace \,. \label{num2} \eea

The first term on the r.h.s.~of Eq.~(\ref{num2}) gives the Sivers
contribution while the second term gives the transversity $\otimes$
Collins effect. We have numerically checked that the third term
gives negligible contributions. Notice that the various terms of the
type $\sin(\phi_S - \Phi)$ appearing in Eq.~(\ref{num2}) can be
decomposed as $\sin\phi_S \cos\Phi - \cos\phi_S \sin\Phi$: similarly
to what has been explicitly shown in appendix~\ref{kin-jet} [see the comment
after Eq.~(\ref{xmin})], the $\cos\phi_S$ terms integrate to zero.
Thus, one obtains the simple expression of Eq.~(\ref{ds1}), given
for $\phi_S = \pi/2$.
%

%
%
\end{document}